\documentclass[aps,prb,twocolumn,showpacs,groupedaddress]{revtex4}

\usepackage[T1]{fontenc}
\usepackage[english]{babel}
\usepackage{epsfig}
\usepackage{graphicx}
\usepackage{color}
\usepackage{amsmath}
\usepackage{bbold}
\usepackage{float}
\usepackage{subfigure}

\newcommand{\bs}[1] {\mathbf{#1}}
\newcommand{\fl} [2]{#1^{\Lambda}_{\mathrm{#2} }  }
\newcommand{\up}{\uparrow}
\newcommand{\down}{\downarrow}

\newcommand{\gWeiss}{\mathcal{G}_{0,{\rm aim}}}

\begin{document}

\title{Antiferromagnetic and d-wave pairing correlations in the strongly interacting two-dimensional Hubbard model from the functional renormalization group}

\author {Demetrio~Vilardi}
\author{Ciro~Taranto}
\author{Walter~Metzner}
\affiliation{Max Planck Institute for Solid State Research, Heisenbergstrasse 1, D-70569 Stuttgart, Germany}

\date{\today}

%
%
%
%

\begin{abstract}
Using the dynamical mean-field theory (DMFT) as a ``booster-rocket'', the functional renormalization group (fRG) can be upgraded from a weak-coupling method to a powerful computation tool for strongly interacting fermion systems.
The strong local correlations are treated non-perturbatively by the DMFT, while the fRG flow can be formulated such that it is driven exclusively by non-local correlations, which are more amenable to approximations.
We show that the full frequency dependence of the two-particle vertex needs to be taken into account in this approach, and demonstrate that this is actually possible -- in spite of the singular frequency dependence of the vertex at strong coupling.
We are thus able to present the first results obtained from the DMFT-boosted fRG for the two-dimensional Hubbard model in the strongly interacting regime.
We find strong antiferromagnetic correlations from half-filling to 18 percent hole-doping, and, at the lowest temperature we can access, a sizable $d$-wave pairing interaction driven by magnetic correlations at the edge of the antiferromagnetic regime.
\end{abstract}

\pacs{}
\maketitle


\section{Introduction}

The discovery of high temperature superconductivity in cuprates raised new challenges in the field of strongly correlated electron systems. \cite{Keimer2015}
Anderson \cite{Anderson1987} proposed the two-dimensional single-band Hubbard model to describe the electron dynamics in the copper-oxide planes.
While the Hubbard model cannot be expected to capture all aspects of cuprate superconductors, it describes their most important property, that is, $d$-wave superconductivity close to an antiferromagnetic order. \cite{Scalapino2012}
In spite of the apparent simplicity of the Hubbard Hamiltonian, a solution of the model in the strong coupling regime relevant for cuprate superconductors turned out to be extremely difficult.

For weak and moderate interactions, the functional renormalization group (fRG) provided conclusive evidence for $d$-wave superconductivity in the two-dimensional Hubbard model. \cite{Metzner2012} With its unbiased treatment of all fluctuation channels on equal footing, the fRG confirmed earlier studies of $d$-wave pairing based on the summation of certain perturbative contributions. \cite{Scalapino2012}

The fRG is based on an exact hierarchy of flow equations for the effective interactions and the self-energy of the system. \cite{Metzner2012,Wetterich1993} Truncations of this hierarchy are possible for weak or moderate interactions, but not at strong coupling. In particular, the truncated fRG equations used so far do not capture the Mott transition, which plays a crucial role in the strongly interacting Hubbard model.

The Mott metal-insulator transition in the Hubbard model is essentially a consequence of strong {\em local}\/ correlations.
As such, it is well described by the dynamical mean field theory (DMFT), \cite{Metzner1989,Georges1991,Georges1996} which treats local correlations non-perturbatively. The DMFT is exact in the limit of infinite dimensions, where non-local correlations are absent. \cite{Metzner1989,Georges1996}
The single-site DMFT has been extended to self-consistent cluster approximations to take also short-ranged non-local correlations into account. \cite{Maier2005}
Long-ranged correlations have been added to the DMFT solution by several perturbative methods. \cite{Rohringer2017,Toschi2007}

Recently, Taranto \emph{et al.} \cite{Taranto2014} managed to combine the strengths of the DMFT and the fRG in a new computational method, the DMF$^2$RG. In this approach the fRG flow does not start from the bare action of the system, but rather from the DMFT solution. Local correlations are thus included already from the beginning, and non-local correlations are generated by the fRG flow. In particular, the Mott physics at strong coupling is captured via the DMFT starting point. The weaker non-local correlations may be captured by a managable truncation of the exact fRG hierarchy.

A key object in the fRG flow is the two-particle vertex, since it determines the two-particle correlations and effective interactions, and the flow of the self-energy.
In a translation invariant system, the two-particle vertex is a function of three independent momentum and frequency variables. A suitable parametrization of these complicated dependences is difficult. 
Taranto \emph{et al.} \cite{Taranto2014} used a channel decomposition \cite{Karrasch2008,Husemann2009,Husemann2012} to reduce the frequency dependence to one frequency variable in each channel, and the momentum dependence was discretized by a rough partition of the Brillouin zone. These approximations limited the application of the DMF$^2$RG to the weak-to-moderate coupling range. Momentum dependences of the two-particle vertex and the self-energy were computed from the DMF$^2$RG flow at half-filling for moderate coupling strengths. \cite{Taranto2014}

For strong interactions, the two-particle vertex exhibits strong frequency dependences which cannot be reduced to one frequency per interaction channel. This is obvious already at the DMFT level. \cite{Rohringer2012}
The full frequency dependence of the vertex is required to compute response functions within the DMFT. \cite{Toschi2012,Hafermann2014,Vilardi2018}
In a recent fRG study it was shown that non-separable frequency dependences are generated even for moderate interactions. \cite{Vilardi2017}

The application of the DMF$^2$RG at strong coupling thus requires an accurate parametrization of the full frequency dependence of the vertex. This is extremely challenging, since the frequency dependence is not only complicated, but also singular at strong coupling. We have overcome these difficulties by various technical developments, so that we are now able to compute the first DMF$^2$RG flows for the strongly interacting two-dimensional Hubbard model. We will present results both at half-filling, where N\'eel antiferromagnetism is the only physical instability, and away from half-filling, where $d$-wave pairing emerges.

Sec.~II is dedicated to methodological aspects. Here we describe the flow equations and our parametrization of the two-particle vertex. In addition to a more accurate parametrization of the vertex, a major advance compared to the first version of the DMF$^2$RG is a setup of the flow that conserves local correlations (already captured by the DMFT). In other words, only non-local correlations are generated by the flow. This substantially improves the accuracy of the unavoidable truncation of the flow equation hierarchy.

In Sec.~III we present results obtained from the DMF$^2$RG for the two-dimensional Hubbard model at strong coupling, in the regime that applies to cuprates. The fRG hierarchy is truncated at the two-particle level, that is, only the influence of non-local three-particle interactions (and beyond) is neglected.
The frequency dependence of the two-particle vertex and the self-energy is fully taken into account. The momentum dependence of the two-particle vertex is approximated by $s$-wave and $d$-wave form factors.
Due to the unbiased treatment of all two-particle interaction channels, we capture the complete interplay of charge, magnetic, and pairing fluctuations.
All calculations are carried out at finite temperature; the lowest temperatures reached are two orders of magnitude smaller than the band width.
Antiferromagnetic fluctuations dominate over a wide doping range. They are of N\'eel type at half-filling, but incommensurate for a sizable doping. Strong $d$-wave pairing correlations emerge at the edge of the antiferromagnetic regime. For the lowest temperature we can reach, the model is very close to a superconducting instability.
The pairing mechanism is clearly magnetic, similar to the mechanism at weak coupling as seen in the plain fRG. \cite{Metzner2012}

In Sec.~IV we conclude with a summary and ideas on further developments.


\section{Formalism}


\subsection{Model}

The Hubbard model\cite{Montorsi1992} describes spin-$\frac{1}{2}$ lattice fermions with inter-site hopping amplitudes $t_{ij}$ and a local interaction $U$. The Hamiltonian is given by
\begin{equation}
 \mathcal{H} = \sum_{i,j,\sigma} t_{ij} c^{\dagger}_{i,\sigma} c_{j,\sigma}
 + U \sum_{i} n_{i,\uparrow} n_{i,\downarrow} ,
\end{equation}
where $c^{\dagger}_{i,\sigma}$ ($c_{i,\sigma}$) creates (annihilates) fermions on site $i$ with spin orientation $\sigma$ ($\uparrow$ or $\downarrow$), and
$n_{i,\sigma} = c^{\dagger}_{i,\sigma} c_{i,\sigma}$. We consider the two-dimensional case on a square lattice and repulsive interaction $U>0$ at finite temperature $T$. The hopping amplitude is restricted to $t_{ij} = -t$ for nearest neighbors and $t_{ij} = -t'$ for next-to-nearest neighbors. Fourier transforming the hopping matrix yields the bare dispersion relation
\begin{equation}
 \varepsilon_{\mathbf{k}} =
 -2t \left( \cos{k_x} + \cos{k_y} \right) -4 t' \cos{k_x} \cos{k_y} .
\end{equation}

Both DMFT and fRG are formulated in a functional integral formalism. The bare action corresponding to the Hubbard Hamiltonian has the form
\begin{align}
 {\cal S} = & - \int_0^{\beta} d\tau d\tau' \sum_{\bs{k},\sigma} 
 \bar{\psi}_{\bs{k},\sigma}(\tau) 
 G_0^{-1}(\bs{k},\tau-\tau') \psi_{\bs{k},\sigma}(\tau')
\nonumber \\
 & + U \int_0^{\beta} d\tau \sum_i n_{i,\up} (\tau) n_{i,\down} (\tau) ,
\label{eq:bareAction}
\end{align}
with $n_{i,\sigma}(\tau) = \bar\psi_{i,\sigma}(\tau) \psi_{i,\sigma}(\tau)$.
Here $\bar\psi_{i,\sigma}(\tau)$ and $\psi_{i,\sigma}(\tau)$ are imaginary time Grassmann fields corresponding to the creation and annihilation operators $c^{\dagger}_{i,\sigma}$  and $c_{i,\sigma}$, respectively, while $\bar\psi_{\bs{k},\sigma}(\tau)$ and $\psi_{\bs{k},\sigma}(\tau')$ are their Fourier components in momentum space.
The kernel of the quadratic part of $\cal S$ is the inverse bare imaginary time propagator. In Matsubara frequency representation it has the simple form
$G_0^{-1}(\bs{k},\nu) = i\nu + \mu  - \epsilon_{\bs{k}}$, where $\mu$ is the chemical potential.


\subsection{DMF$\bf^2$RG}

The DMFT treats local correlations non-perturba\-tive\-ly, while non-local correlations are neglected. \cite{Georges1996} It is exact in the limit of infinite lattice dimensions, where non-local correlations vanish. \cite{Metzner1989}
In the absence of non-local correlations, the self-energy is local and a functional of the local propagator. Hence, the lattice problem can be mapped to a single Hubbard site coupled to a non-interacting fermionic bath, that is, to an auxiliary single impurity Anderson model. \cite{Georges1991} The self-energies and local propagators of the impurity and the lattice model must coincide, which leads to the DMFT self-consistency condition
\begin{align}
 G_{\rm loc}(\nu) & = 
 \int_{\bs{k}} \frac{1}{i\nu + \mu - \epsilon_{\bs{k}} - \Sigma_{\rm dmft}(\nu)}
 \nonumber \\ &= 
 \frac{1}{\gWeiss^{-1}(\nu) - \Sigma_{\rm dmft}(\nu)} \, .
\label{eq:DMFTloop}
\end{align}
Here and in the following $\int_{\bs{k}}$ is a short-hand notation for $\int \frac{d^2\bs{k}}{(2\pi)^2}$, and $\gWeiss$ is the bare propagator of the Anderson impurity model (AIM). The self-energy $\Sigma_{\rm dmft}$ can be obtained by solving the AIM with $\gWeiss$ and $U$.

The fRG is based on a scale-by-scale evaluation of the many-body functional integral. \cite{Metzner2012}
A flow is generated by letting the propagator in the quadratic part of the bare action depend on a flow parameter $\Lambda$. For the Hubbard model, this leads to an action of the form
\begin{align}
 {\cal S}^{\Lambda} = & 
 - \int_0^{\beta} d\tau d\tau' \sum_{\bs{k},\sigma} \bar{\psi}_{\bs{k},\sigma}(\tau) 
 {G^{\Lambda}_0}^{-1}(\bs{k},\tau-\tau') \psi_{\bs{k},\sigma}(\tau')
 \nonumber \\
 & + U \int_0^{\beta} d\tau \sum_i n_{i,\up}(\tau) n_{i,\down}(\tau) \, .
\label{eq:regAction}
\end{align}
The scale dependence of the function $G^{\Lambda}_0$ generates a flow for the generating functionals. The flow of the generating functional for one-particle irreducible (1PI) vertex functions, the effective action $\Gamma^{\Lambda}[\psi,\bar\psi]$, is governed by an exact functional flow equation. \cite{Wetterich1993}
The propagator $G^{\Lambda_{\rm ini}}_0$ at the initial value $\Lambda=\Lambda_{\rm ini}$ of the flow parameter determines the initial condition of the flow. The final result at $\Lambda_{\rm fin}$ is determined by the condition 
$G^{\Lambda_{\rm fin}}_0 = G_0 = \left( i\omega + \mu - \epsilon_{\bs{k}}\right)^{-1}$, 
restoring the original action (\ref{eq:bareAction}).
The initial condition for the effective action 
$\Gamma_{\rm ini}[\psi,\bar{\psi}] = \Gamma^{\Lambda_{\rm ini}} [\psi,\bar{\psi}]$ 
is determined by the function $G_0^{\Lambda_{\rm ini}}$.
In the conventional fRG, $G_0^{\Lambda_{\rm ini}} = 0$ is chosen, \cite{Metzner2012}
leading to an uncorrelated starting point. 

In order to start from the DMFT solution, we impose \cite{Taranto2014}
\begin{equation}
G^{\Lambda_{\rm ini}}_0(\bs{k},\tau-\tau') = \gWeiss(\tau-\tau'),
\label{eq:G0LInit}
\end{equation}
where $\gWeiss$ is the bare propagator of the AIM fulfilling the self-consistency relation (\ref{eq:DMFTloop}). 
In this way the intial value for the action~(\ref{eq:regAction}) becomes
\begin{align}
 {\cal S}^{\Lambda_{\rm ini}} = &
 - \int_0^{\beta} d\tau d\tau' \sum_{\bs{k},\sigma} 
 \bar{\psi}_{\bs{k},\sigma}(\tau) \gWeiss^{-1}(\tau-\tau') \psi_{\bs{k},\sigma}(\tau')
 \nonumber \\
 & + U \int_0^{\beta} \sum_i d\tau n_{i,\up}(\tau) n_{i,\down}(\tau) \, .
\label{eq:AIMAction}
\end{align}
%
Action~(\ref{eq:AIMAction}) determines the initial condition for $\Gamma^{\Lambda}[\psi,\bar{\psi}]$ as
\begin{equation}
\Gamma_{\rm ini}[\psi,\bar{\psi}] = \Gamma_{\rm dmft}[\psi,\bar{\psi}]. 
\label{eq:initDMF2RGAction}
\end{equation}
Hence, the initial condition of the functional flow is given by the effective action of the self-consistent Anderson impurity model.

By expanding the exact flow equation in powers of the Grassmann fields, one obtains an infinite hierarchy of flow equations for 1PI functions. \cite{Metzner2012}
In the following we will truncate this hierarchy at the two-particle level, that is, we neglect the influence of the three-particle vertex. The flow thus involves only the 
self-energy $\Sigma^{\Lambda}$ and the two-particle vertex $V^{\Lambda}$.
Since local correlations are treated non-perturbatively by the DMFT starting point, the truncation of the flow affects only non-local correlations, and the feedback of non-local correlations on local correlations.

To summarize, the DMF$^2$RG is composed of two steps. First, we solve the DMFT self-consistency loop leading to the local DMFT self-energy $\Sigma_{\rm dmft}$. 
From the AIM with the self-consistent propagator ${\cal G}_{0,{\rm aim}}$ we also compute the local vertex of the AIM, $V_{\rm dmft} = V_{\rm aim}$. 
In a second step, we use the fRG flow equations for the self-energy and the vertex with the local initial conditions $\Sigma_{\rm dmft}$ and $V_{\rm dmft}$, respectively. 
The flow equations for $\Sigma^\Lambda$ and $V^\Lambda$ are formally identical to the 
conventional ones; the only difference is the non-trivial initial condition.
We finally note that in the DMF$^2$RG the ``scale'' $\Lambda$ is not a simple energy or momentum scale as in the conventional fRG, but rather a parameter tuning the amount of non-local correlations in the system.


\subsection{Truncated flow equations}

\begin{figure}[t!]
\begin{center}
\vskip 3mm
\includegraphics[width=0.3\textwidth]{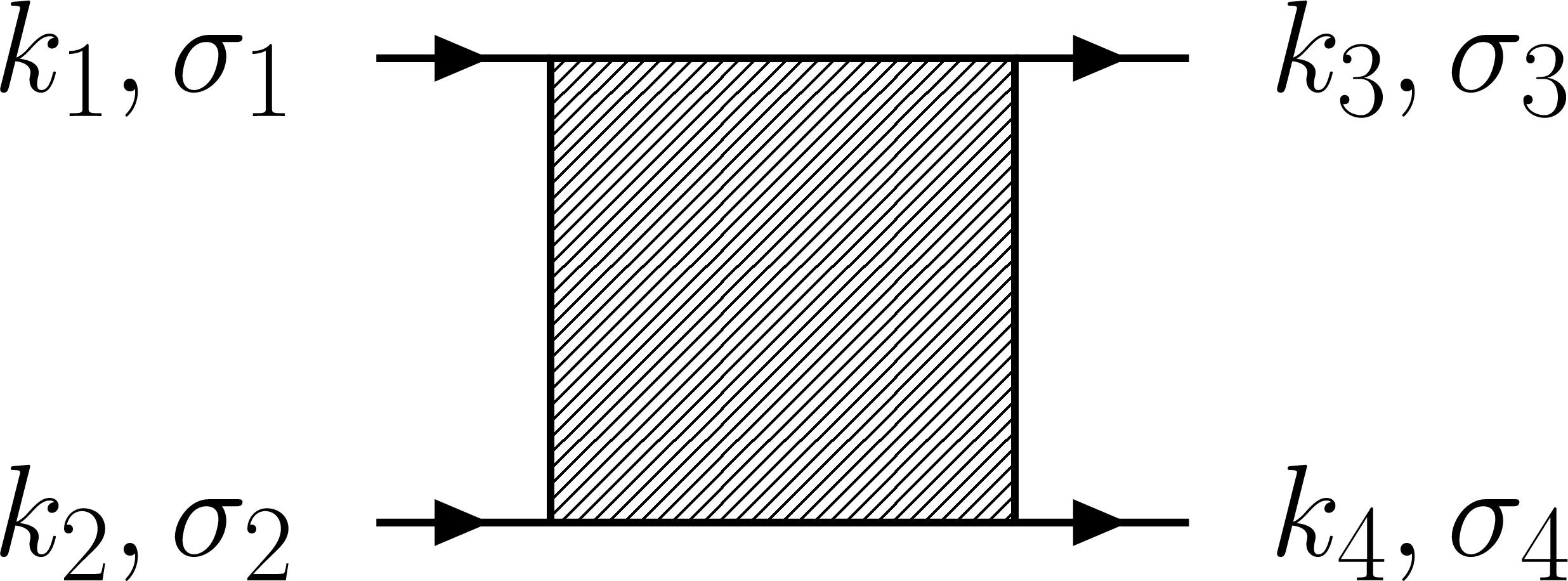}
\caption{Notation of the two-particle vertex.}
\label{fig:notvert} 
\end{center}
\end{figure}
The vertex $V^\Lambda$ depends on four frequency-momentum variables $k_1,\dots,k_4$ and on four spin indices $\sigma_1,\dots,\sigma_4$, as illustrated diagrammatically in Fig.~\ref{fig:notvert}. Here and in the following, momentum and frequency variables are collected in one symbol as $k_i = (\mathbf{k}_i,\nu_i)$. Momentum and energy conservation implies that the vertex can be written as a function of only three variables $k_1,k_2,k_3$.

Due to SU$(2)$ spin symmetry, the self-energy is diagonal in spin space, while all the six non-zero components of the vertex can be expressed in terms of one function \cite{Rohringer2012,Vilardi2017}
$V^{\Lambda}(k_1,k_2,k_3) = V^{\Lambda}_{\up\down\up\down}(k_1,k_2,k_3)$ through the five relations
$V^\Lambda_{\uparrow\uparrow\uparrow\uparrow} = V^\Lambda_{\downarrow\downarrow\downarrow\downarrow}$, 
$V^\Lambda_{\uparrow\downarrow\uparrow\downarrow} = V^\Lambda_{\downarrow\uparrow\downarrow\uparrow}$,
$V^\Lambda_{\uparrow\downarrow\downarrow\uparrow } = V^\Lambda_{\downarrow\uparrow\uparrow\downarrow}$,
\begin{eqnarray}
V^\Lambda_{\uparrow\uparrow\uparrow\uparrow}(k_1,k_2,k_3) &=& V^\Lambda_{\uparrow\downarrow\uparrow\downarrow}(k_1,k_2,k_3) 
\nonumber \\ 
&-& V^\Lambda_{\uparrow\downarrow\uparrow\downarrow}(k_1,k_2,k_1+k_2-k_3) ,
\label{eq:spinsym1}
\\ 
V^\Lambda_{\uparrow\downarrow\downarrow\uparrow}(k_1,k_2,k_3)& =& -V^\Lambda_{\uparrow\downarrow\uparrow\downarrow}(k_1,k_2,k_1+k_2-k_3) . \hskip 5mm
\label{eq:spinsym2}
\end{eqnarray}

The flow equation for the self-energy has the form \cite{Metzner2012}
\begin{equation}
 \frac{d}{d \Lambda} \Sigma^\Lambda(k) =
 \int_p  S^\Lambda(p)\left[2V^\Lambda(k,p,p) -V^\Lambda(k,p,k)\right] \, , 
\label{eq:selfflow}
\end{equation}
where $\int_p = T \sum_\omega \int_{\mathbf{p}}$ is a short-hand notation for the Matsubara frequency sum and the momentum integration over the first Brillouin zone. 
\begin{equation}
 S^\Lambda = \left.
 \frac{dG^\Lambda}{d\Lambda}\right|_{\Sigma^{\Lambda}=\mathrm{const}} 
\end{equation}
is the so-called single-scale propagator, while ${G^\Lambda}$ is the full propagator, which is related to the bare propagator and the self-energy by the Dyson equation
$(G^\Lambda)^{-1} = (G_0^\Lambda)^{-1} - \Sigma^\Lambda$. 
  
The flow equation for the two-particle vertex can be written as \cite{Metzner2012, Husemann2009}
\begin{widetext}
\begin{eqnarray}
\label{eq:vertflow}
 \frac{d}{d\Lambda}V^\Lambda(k_1,k_2,k_3) &=&
 \fl{\mathcal{T}}{pp}(k_1,k_2,k_3) + \fl{\mathcal{T}}{ph}(k_1,k_2,k_3)
 + \fl{\mathcal{T}}{phc}(k_1,k_2,k_3) ,
\end{eqnarray} 
where
%
\begin{eqnarray}
\label{eq:ppT} 
\fl{\mathcal{T}}{pp}(k_1,k_2,k_3) &=& - \int_p \fl{\mathcal{P}}{\mathrm{pp}}(k_1+k_2,p) \fl{V}{}(k_1,k_2,k_1+k_2-p)\fl{V}{}(k_1+k_2-p,p,k_3) ,
\label{eq:tpp} 
\\ 
\label{eq:tph} 
\fl{\mathcal{T} } {ph}(k_1,k_2,k_3) & =& \int_p \fl{\mathcal{P}}{\mathrm{ph}}(k_3-k_1,p)
\Big\{ 2 \fl{V}{}( k_1,k_3-k_1+p,k_3)  \fl{V}{}(p,k_2,k_3-k_1+p) \\
\nonumber
&&- \fl{V}{}( k_1,k_3-k_1+p,p)  \fl{V}{}(p,k_2,k_3-k_1+p) - \fl{V}{}( k_1,k_3-k_1+p,k_3)  \fl{V}{}(k_2,p,k_3-k_1+p) \Big\} , \\
\label{eq:tphc}
\fl{\mathcal{T}}{phc}(k_1,k_2,k_3) & =& -\int_p \fl{\mathcal{P}}{\mathrm{ph}}(k_2-k_3,p) \fl{V}{}(k_1,k_2-k_3+p,p)
\fl{V}{}(p,k_2,k_3).
\end{eqnarray}
\end{widetext}
Here $\mathcal{T}^\Lambda_{\mathrm{pp}}$, $\mathcal{T}^\Lambda_{\mathrm{ph}}$ and $\mathcal{T}^\Lambda_{\mathrm{phc}}$ stand, respectively, for \textit{particle-particle}, \textit{particle-hole} and \textit{particle-hole crossed} contributions.
We have defined the quantities
\begin{align}
 \mathcal{P}_{\mathrm{ph}}^\Lambda(Q,p) &= G^\Lambda(Q+p)S^\Lambda(p) + G^\Lambda(p) S^\Lambda(Q+p), \\ 
 \mathcal{P}_{\mathrm{pp}}^\Lambda(Q,p) &=G^\Lambda(Q-p)S^\Lambda(p) + G^\Lambda(p) S^\Lambda(Q-p) ,
\end{align}
which are the scale derivatives, at fixed self-energy, of the product of two Green's functions.

The initial conditions for the flow equations (\ref{eq:selfflow}) and (\ref{eq:vertflow}) are
\begin{eqnarray}
  \Sigma^{\Lambda_{\rm ini}}(\mathbf{k},\nu) &=& \Sigma_{\rm dmft}(\nu) , \\
  V^{\Lambda_{\rm ini}}(k_1,k_2,k_3) &=& V_{\rm dmft}(\nu_1,\nu_2,\nu_3) ,
\label{eq:initvert}
\end{eqnarray}
where $\nu_i$ is the frequency component of $k_i = (\mathbf{k}_i,\nu_i)$.


\subsection{DMFT conserving flow}
\label{sec:cutoff}

There is much freedom in choosing the $\Lambda$ dependence of the bare propagator 
$G^{\Lambda}_0$, which can be exploited to our advantage. The initial condition is determined by Eq.~(\ref{eq:G0LInit}), while the final value is given by the bare lattice propagator, $G^{\Lambda_{\rm fin}}_0 = G_0$. Taranto \emph{et al.} \cite{Taranto2014} used a linear interpolation between the initial and final values.

Here, we choose $G_0^{\Lambda}$ such that 
\begin{equation}
  \left. G^{\Lambda}_{\rm loc}(\nu) \right|_{\Sigma^{\Lambda} = \Sigma_{\rm dmft}} =
  \int_{\bs{k}} 
  \left. G^{\Lambda}(\bs{k},\nu) \right|_{\Sigma^{\Lambda} = \Sigma_{\rm dmft}}
  = G_{\rm dmft}(\nu) ,
\label{eq:localCutoffCondition}
\end{equation}
is independent of $\Lambda$, and thus given by the local propagator as obtained from the DMFT. This can be achieved by an ansatz of the form
\begin{equation}
 G^{\Lambda}_{0}(\bs{k},\nu) =
 \frac{1}{ i\nu + \mu - (1 - \Lambda) \epsilon_{\bs{k}} - g^{\Lambda}(\nu) \Delta(\nu)},
\label{eq:locCutoff}
\end{equation}
with the hybridization function $\Delta(\nu) = i\nu + \mu - \gWeiss^{-1}(\nu)$, and a function $g^{\Lambda}(\nu)$ which is determined by the condition (\ref{eq:localCutoffCondition}).
The initial value for $g^{\Lambda}(\nu)$ is $g^{\Lambda_{\rm ini}}(\nu) = 1$ at $\Lambda_{\rm ini} = 1$, such that the condition~(\ref{eq:G0LInit}) is fulfilled. 
The final value is $g^{\Lambda_{\rm fin}}(\nu) = 0$ at $\Lambda_{\rm fin} = 0$. 
The value of the chemical potential $\mu$ is fixed and determined by the DMFT solution. 
The simple choice $g^{\Lambda}(\nu) = \Lambda$ yields the flow scheme used in Ref.~\onlinecite{Taranto2014}, where $G_{\rm loc}^\Lambda$ is scale dependent even if $\Sigma^\Lambda = \Sigma_{\rm dmft}$ is kept fixed.

Inserting the ansatz (\ref{eq:locCutoff}) into the condition (\ref{eq:localCutoffCondition}), we obtain the equation
\begin{align}
 & \int_{\bs{k}} \frac{1} {i\nu + \mu - (1-\Lambda)\epsilon_{\bs{k}} 
 - g^{\Lambda}(\nu) \Delta(\nu) - \Sigma_{\rm dmft}(\nu) }
 \nonumber \\[1mm]
 & = \big[ \gWeiss^{-1}(\nu) - \Sigma_{\rm dmft}(\nu) \big]^{-1} ,
\end{align}
from which we can determine $g^{\Lambda}(\nu)$ numerically for any $\Lambda$. 
\begin{figure}
 \includegraphics[width=0.48\textwidth]{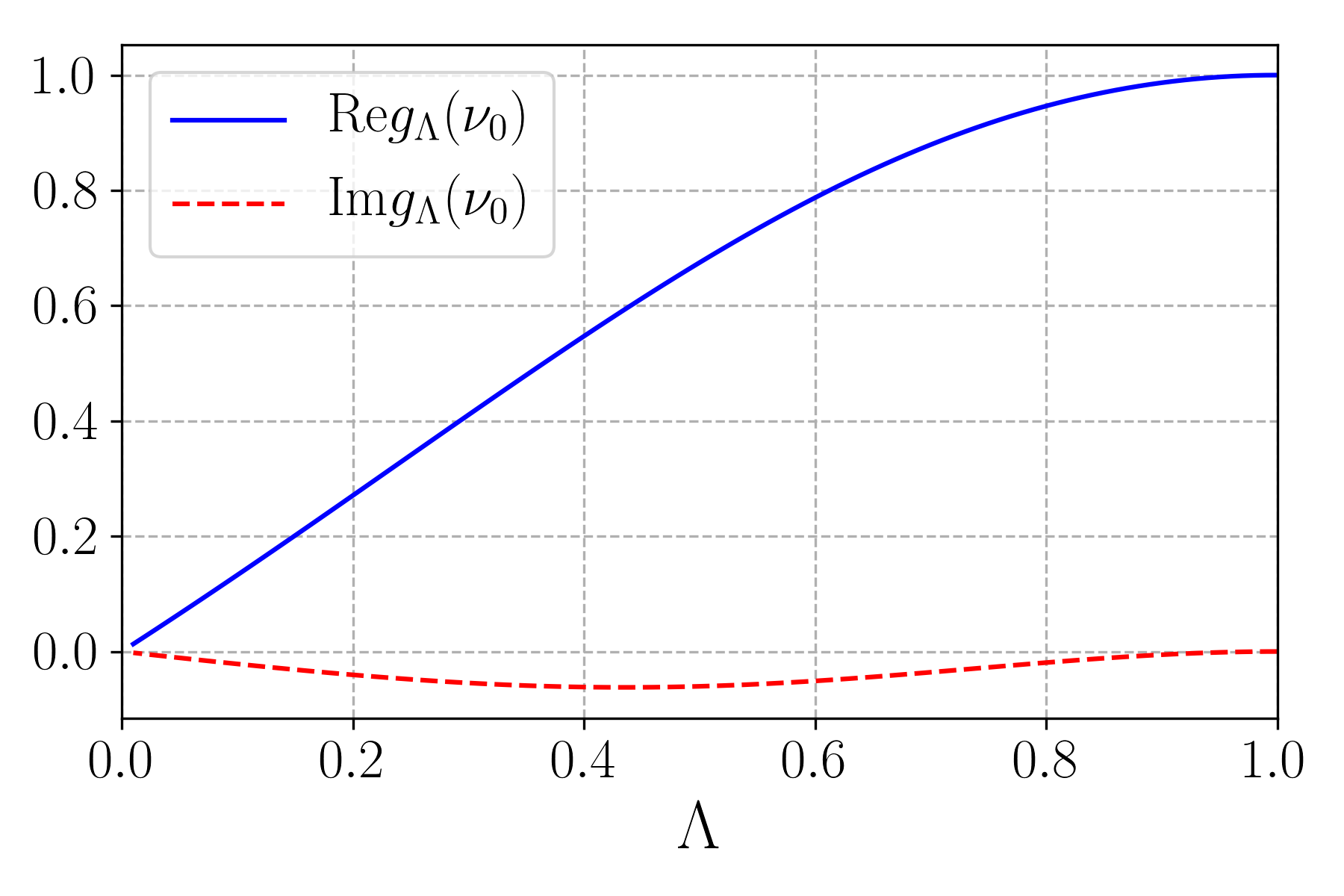}
 \caption{$g^{\Lambda}(\nu)$ as defined in Eq.~(\ref{eq:locCutoff}) as a function 
 of $\Lambda$ for the first Matsubara frequency $\nu_0 = \pi T$. 
 Parameters are $n=0.82$, $U=8t$, $T=0.08t$, and $t'=-0.2t$.}
\label{fig:DMFTConservingCutoff} 
\end{figure}
In Fig.~\ref{fig:DMFTConservingCutoff}, we show an example for the function $g^{\Lambda}(\nu)$ as a function of $\Lambda$ for the first Matsubara frequency $\nu_0 = \pi T$.
In the absence of particle-hole symmetry, $g^\Lambda(\nu)$ has a non-zero imaginary part. The real part is linear in $\Lambda$ only for small $\Lambda$. The frequency dependence of $g^{\Lambda}(\nu)$ (not shown here) is very weak.

From Eq.~(\ref{eq:locCutoff}), we can calculate the single-scale propagator
\begin{equation}
  S^{\Lambda} = - G^{\Lambda} \frac{d(G^{\Lambda}_0)^{-1}}{d\Lambda} G^{\Lambda} =
  -G^{\Lambda} \left[ \epsilon_{\bs{k}} - \Delta \frac{d g^{\Lambda}}{d\Lambda} \right] G^{\Lambda} .
\end{equation}
The function $dg^{\Lambda}/d\Lambda$ can be conveniently determined by taking the 
$\Lambda$-derivative of Eq.~(\ref{eq:localCutoffCondition}).

\begin{figure}
 \includegraphics[width=0.2\textwidth]{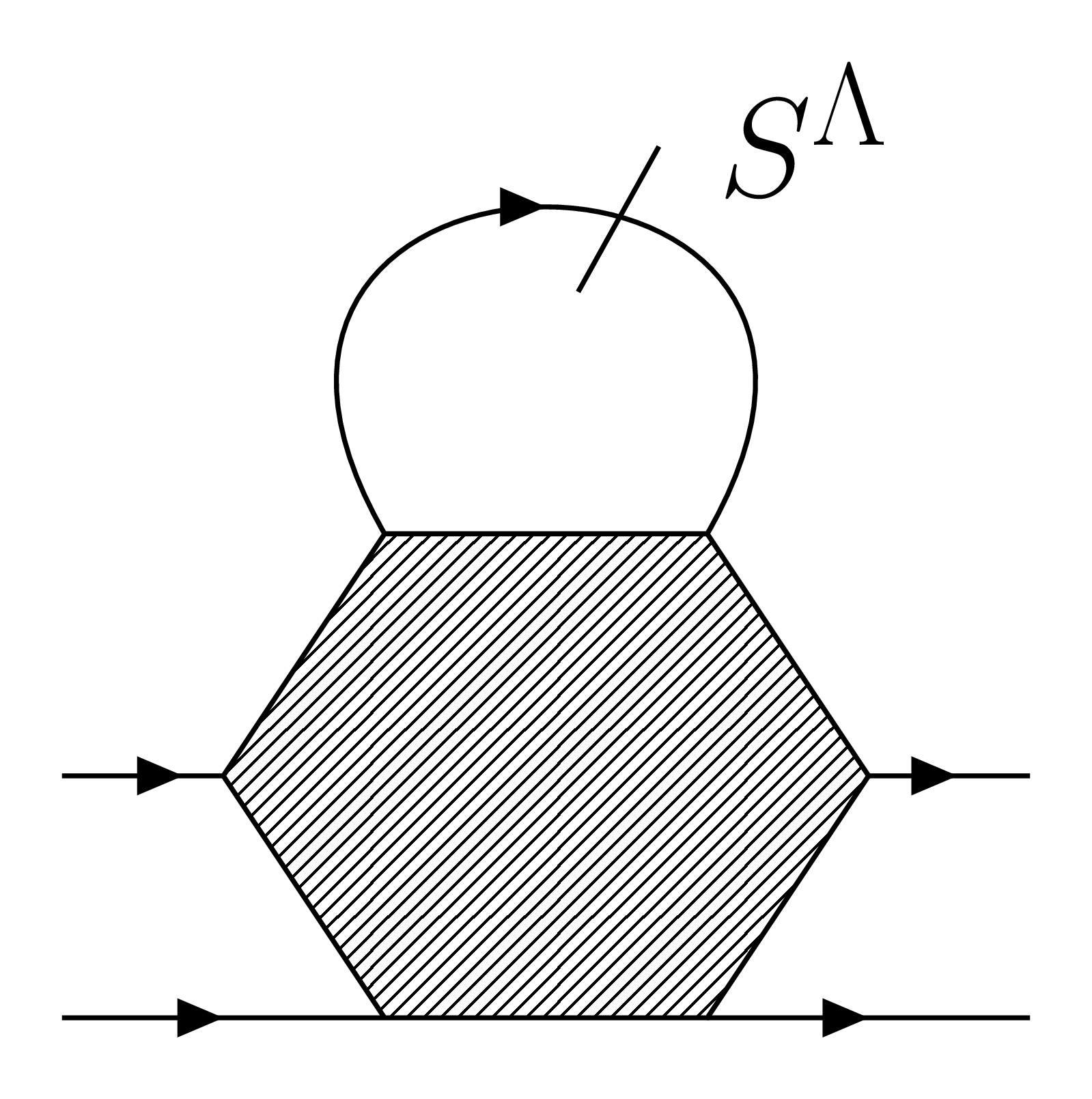}
 \caption{Contribution from three-particle vertex to the flow of the two-particle vertex.}
\label{fig:3p-tadpole} 
\end{figure}

The condition (\ref{eq:localCutoffCondition}) implies that the local single-scale propagator $S_{\rm loc}^{\Lambda}$ with $\Sigma^\Lambda = \Sigma_{\rm dmft}$ vanishes.
Hence, for $\Sigma^\Lambda = \Sigma_{\rm dmft}$, there are no local contributions to the flow, and thus no corrections to the DMFT solution. In other words, the DMFT is conserved by the flow. The flow is thus exclusively generated by non-local contributions.
This improves the accuracy of the truncation. In particular, the three-particle contributions to the flow of the two-particle vertex (see Fig.~\ref{fig:3p-tadpole}), which are neglected in our truncation, contribute only after non-local correlations have been generated. At the initial stage of the flow, where the three-particle vertex is local and the self-energy is given by $\Sigma_{\rm dmft}$, the three-particle tadpole contribution to the flow of $\Gamma^{\Lambda}$ vanishes.
%


\subsection{Vertex parametrization} 
\label{sec:vertex}

We parametrize the two-particle vertex by extending the channel decomposition introduced by Husemann and Salmhofer. \cite{Husemann2009,Husemann2012} This allows us to capture the dominant momentum dependence generated by each interaction channel by an accurate discretization, while the remaining more regular momentum dependences are approximated by a few form factors.
The vertex function $V^{\Lambda}(k_1,k_2,k_3)$ is decomposed as
\begin{eqnarray}
 V^{\Lambda}(k_1,k_2,k_3)&=& V_{\rm dmft}(\nu_1,\nu_2,\nu_3) - \phi^{\Lambda}_{\mathrm{p}}(k_1+k_2;k_1,k_3) \nonumber \\
 &+& \phi^{\Lambda}_{\mathrm{m}}(k_2-k_3;k_1,k_2) \nonumber \\ &+&
 {\textstyle \frac{1}{2}}  \phi^{\Lambda}_{\mathrm{m}}(k_3-k_1;k_1,k_2) \nonumber \\ 
 &-& {\textstyle \frac{1}{2}} \phi^{\Lambda}_{\mathrm{c}}(k_3-k_1;k_1,k_2),
\label{eq:decomposition}
\end{eqnarray}
with the DMFT vertex $V_{\rm dmft}$, the \emph{pairing} channel $\phi_{\rm p}^\Lambda$, the \emph{magnetic} channel $\phi_{\rm m}^\Lambda$, and the \emph{charge} channel $\phi_{\rm c}^\Lambda$. 
The functions $\phi_{\rm x}$ in~(\ref{eq:decomposition}) are nonlocal fluctuation contributions
beyond the DMFT solution. Local pairing, magnetic and charge fluctuations are already captured by the DMFT vertex. 
The initial condition for the vertex~(\ref{eq:initvert}) determines the starting conditions $\phi^{\Lambda_{\rm ini}}_{\rm p} = \phi^{\Lambda_{\rm ini}}_{\rm m} = \phi^{\Lambda_{\rm ini}}_{\rm c} = 0$.

To derive the flow equations for the non-local fluctuation channels,  
we substitute Eq.~(\ref{eq:decomposition}) into Eq.~(\ref{eq:vertflow}) and, 
by following Ref.~\onlinecite{Vilardi2017}, we obtain the equations for $\phi_x$
\begin{eqnarray}
\label{eq:phi_p}
 \dot{\phi}_{\mathrm{p}}^{\Lambda}(Q;k_1,k_3) &=&
 -\mathcal{T}^{\Lambda}_{\mathrm{pp}}(k_1,Q-k_1,k_3) , \\
\label{eq:phi_c}
 \dot{\phi}_{\mathrm{c}}^{\Lambda}(Q;k_1,k_2) &=&
 \mathcal{T}^{\Lambda}_{\mathrm{phc}}(k_1,k_2,k_2-Q) \nonumber \\
 && -2\mathcal{T}^{\Lambda}_{\mathrm{ph}}(k_1,k_2,Q+k_1), \\
\label{eq:phi_m}
\dot{\phi}_{\mathrm{m}}^{\Lambda}(Q;k_1,k_2) &=& \mathcal{T}^{\Lambda}_{\mathrm{phc}}(k_1,k_2,k_2-Q) .
\end{eqnarray}

While capturing the full frequency dependence and the dependence on the bosonic momentum $\bs{Q}$ for each channel by an accurate discretization, we treat the remaining fermionic momentum dependence approximately by using a small orthonormal set of form factors $f_l(\bs{k})$. \cite{Vilardi2017}
For the charge and magnetic channels we keep only $f_s(\bs{k}) = 1$, while for the pairing channel we use $f_s(\bs{k}) = 1$ and $f_d(\bs{k}) = \cos k_x - \cos k_y$.
This is sufficient to capture the leading magnetic and pairing instabilities.
The non-local fluctuation terms can thus be written as
\begin{align}
\label{eq:SD}
 \phi^{\Lambda}_{\mathrm{p}}(Q;k_1,k_3) & =
 \mathcal{S}^\Lambda_{\bs{Q},\Omega}(\nu_1,\nu_3) \nonumber \\ 
 & \hskip -1.5cm 
 + f_d \left(\bs{Q}/2 - \bs{k}_1\right) f_d\left(\bs{Q}/2 - \bs{k}_3\right) \mathcal{D}^\Lambda_{\bs{Q},\Omega}(\nu_1,\nu_3), \\
\label{eq:C}
 \phi^\Lambda_{\mathrm{c}}(Q;k_1,k_2) &= \mathcal{C}^\Lambda_{\bs{Q},\Omega}(\nu_1,\nu_2), \\
\label{eq:M}
 \phi^\Lambda_{\mathrm{m}}(Q;k_1,k_2) &= \mathcal{M}^\Lambda_{\bs{Q},\Omega}(\nu_1,\nu_2),
\end{align}
where the functions ${\cal S}^\Lambda$, ${\cal D}^\Lambda$, ${\cal C}^\Lambda$, and
${\cal M}^\Lambda$ still depend on three frequencies, but only on one momentum variable.

We emphasize that we keep the full frequency dependence in each channel. Below we will show that approximating the channels by functions of a single (bosonic) linear combination of frequencies as in Ref.~\onlinecite{Taranto2014} restricts the application of the DMF$^2$RG to weak or moderate coupling, while the scope of the formalism is to capture strong coupling effects.

The flow equations for $\mathcal{S}^{\Lambda}$, $\mathcal{D}^{\Lambda}$, $\mathcal{C}^{\Lambda}$ and $\mathcal{M}^\Lambda$ are obtained by inserting Eqs.~(\ref{eq:SD}), (\ref{eq:C}) and (\ref{eq:M}) into Eqs.~(\ref{eq:phi_p}), (\ref{eq:phi_c}) and (\ref{eq:phi_m}), respectively, and by projecting onto form factors. \cite{Vilardi2017}
The flow equation for the magnetic channel $\mathcal{M}^{\Lambda}$ reads
\begin{widetext}
\begin{equation}
 \frac{d}{d\Lambda} \mathcal{M}^{\Lambda}_{\bs{Q},\Omega}(\nu_1,\nu_2) = 
 -T \sum_\nu L^{\mathrm{m},\Lambda}_{\mathbf{Q},\Omega}(\nu_1,\nu) 
 P_{\bs{Q},\Omega}^{\Lambda}(\nu) 
 L^{\mathrm{m},\Lambda}_{\mathbf{Q},\Omega}(\nu,\nu_2-\Omega), 
\label{eq:FlowMag}
\end{equation}
with
\begin{equation}
 P_{\bs{Q},\Omega}^{\Lambda}(\omega) = \int_{\bs{p}}
 G^\Lambda(\bs{p},\omega) S^\Lambda(\bs{Q}+\bs{p},\Omega+\omega) +
 G^\Lambda(\bs{Q}+\bs{p},\Omega+\omega) S^\Lambda(\bs{p},\omega) ,
\label{eq:Pph}
\end{equation}
and
\begin{eqnarray} 
\nonumber
 L^{\mathrm{m},\Lambda}_{\bs{Q},\Omega}(\nu_1,\nu_2)
 &=& V_{\rm dmft}(\nu_1, \nu_2, \nu_2 - \Omega) + \mathcal{M}^\Lambda_{\bs{Q},\Omega}(\nu_1,\nu_2) 
\\
 &+& \int_{\bs{p}} \Big \{- \mathcal{S}^\Lambda_{\bs{p},\nu_1+\nu_2}(\nu_1,\nu_1+\Omega)   
 -\frac{1}{2} (\cos Q_x + \cos Q_y) 
 \mathcal{D}^\Lambda_{\bs{p},\nu_1+\nu_2}(\nu_1,\nu_1+\Omega)
\nonumber \\
 &+&\frac{1}{2} \Big[  \mathcal{M}^\Lambda_{\bs{p},\nu_2-\nu_1-\Omega}( \nu_1,\nu_2) 
 - \mathcal{C}^\Lambda_{\bs{p},\nu_2-\nu_1-\Omega}(\nu_1,\nu_2) \Big] 
 \Big \} .
\label{eq:Lxph} 
\end{eqnarray}
The other flow equations are reported in the Appendix.
\end{widetext}


\subsection{Single-channel approximation}
\label{sec:1ch}

In the conventional fRG, when restricting the flow of the two-particle vertex to a single channel, particle-particle or direct/crossed particle-hole, and when neglecting the self-energy feedback, the solution of the flow equation is equivalent to a random phase approximation (RPA) in that particular channel.~\cite{Metzner2012}
A similar statement holds for the DMF$^2$RG: neglecting the self-energy flow, the single-channel DMF$^2$RG is equivalent to a RPA with the irreducible DMFT vertex instead of the bare interaction. The DMFT vertex is required for the calculation of response functions within DMFT. \cite{Georges1996} We now demonstrate this equivalence explicitly for the case of the crossed particle-hole channel. 

Within DMFT, the momentum dependent vertex function for the calculation of magnetic response functions is obtained from the Bethe-Salpeter equation in the crossed particle-hole channel as \cite{Georges1996,Rohringer2012}
\begin{equation}
 V^{\rm rpa}_{\bs{Q},\Omega}(\nu_1,\nu_3) =
 \sum_{\nu} V_{{\rm dmft}, \Omega} (\nu_1,\nu) A^{-1}_{\bs{Q},\Omega}(\nu,\nu_3) , 
\label{eq:ladderDMFT}
\end{equation}
with $V_{{\rm dmft}, \Omega}(\nu_1,\nu_3) = V_{\rm dmft}(\nu_1,\nu_3+\Omega,\nu_3)$,
and
\begin{align}
  A_{\bs{Q},\Omega}&(\nu_1,\nu_3) = \delta_{\nu_1,\nu_3} \nonumber \\
 & - T [\chi^{0}_{\bs{Q},\Omega}(\nu_1)-\chi^{0}_{{\rm loc},\Omega}(\nu_1)]
 V_{{\rm dmft},\Omega}(\nu_1,\nu_3) .
\label{eq:ladderDMFTMatrix}
\end{align}
$A^{-1}_{\bs{Q},\Omega}(\nu_1,\nu_3)$ is the matrix inverse of
$A_{\bs{Q},\Omega}(\nu_1,\nu_3)$ viewed as matrix with the fermionic Matsubara frequencies $\nu_1$ and $\nu_3$ as matrix indices. 
We also introduced the momentum integrated particle-hole propagator
\begin{equation}
\chi^{0}_{\bs{Q},\Omega}(\nu) = - \int_{\bs{k}} G_{\rm dmft}(\bs{k},\nu)G_{\rm dmft}(\bs{Q}+\bs{k},\Omega+\nu) ,
\label{eq:nonlocBubble}
\end{equation}
with
$G_{\rm dmft}^{-1}(\bs{k},\nu) = i\nu +\mu - \epsilon_{\bs{k}} - \Sigma_{\rm dmft}(\nu)$,
and the local particle-hole propagator
$\chi^0_{{\rm loc},\Omega}(\nu) = - G_{\rm loc}(\nu)G_{\rm loc}(\Omega+\nu)$ 
with $G_{\rm loc}(\nu) = \int_{\bs{k}} G_{\rm dmft}(\bs{k},\nu)$. 

To prove the equivalence between DMFT-RPA and single channel DMF$^2$RG, we show that 
Eq.~(\ref{eq:ladderDMFT}) is the solution of the vertex flow equation, once we neglect the flow of the self-energy and take only the crossed particle-hole channel into account. To this end, we introduce the $\Lambda$ dependent particle-hole propagator $\chi^{0,\Lambda}_{\bs{Q},\Omega}(\nu)$ by promoting $G_{\rm dmft}$ in Eq.~(\ref{eq:nonlocBubble}) to the $\Lambda$ dependent propagator 
$G^{\Lambda}_{\rm dmft} = [ {G^{\Lambda}_0}^{-1} - \Sigma_{\rm dmft} ]^{-1}$,
where $G^{\Lambda}_0$ can be any continous function fulfilling the conditions 
$G^{\Lambda_{\rm ini}}_0 = \gWeiss$ and $G^{\Lambda_{\rm fin}}_0 = G_{0,{\rm latt}}$. 
The matrix $A_{\bs{Q},\Omega}$ in Eq.~(\ref{eq:ladderDMFTMatrix}) becomes $\Lambda$ dependent through $\chi^{\Lambda, 0}_{\bs{Q},\Omega}(\nu)$, and Eq.~(\ref{eq:ladderDMFT}) reads
\begin{align}
 V^{{\rm rpa},\Lambda}_{\bs{Q},\Omega}(\nu_1,\nu_3) =
 \sum_{\nu} V_{{\rm dmft},\Omega}(\nu_1,\nu) 
 (A^\Lambda_{\bs{Q},\Omega})^{-1}(\nu,\nu_3) .
\label{eq:ladderDMFTLambda}
\end{align}
Defining the function $\phi^{{\rm rpa},\Lambda} = V^{{\rm rpa},\Lambda} - V_{\rm dmft}$ and taking the $\Lambda$ derivative of Eq.~(\ref{eq:ladderDMFTLambda}) yields
\begin{eqnarray}
 \frac{d\phi^{{\rm rpa},\Lambda}_{\bs{Q},\Omega}}{d\Lambda}  &=&
 T\sum_{\nu} \big[ V_{{\rm dmft},\Omega}(\nu_1,\nu) 
 + \phi^{{\rm rpa},\Lambda}_{\bs{Q},\Omega}(\nu_1,\nu) \big]
\nonumber \\
 &\times& \!\! \frac{d\chi^{\Lambda,0}_{\bs{Q},\Omega}(\nu)}{d\Lambda}
 \big[ V_{{\rm dmft},\Omega}(\nu,\nu_3) 
 + \phi^{{\rm rpa},\Lambda}_{\bs{Q},\Omega}(\nu,\nu_3) \big] . \hskip 6mm
\label{eq:ladderDMFTDot}
\end{eqnarray}
Eq.~(\ref{eq:ladderDMFTDot}) is equivalent to Eq.~(\ref{eq:FlowMag}) with $\mathcal{M}^\Lambda = \phi^{{\rm rpa},\Lambda}$, once the feedback of the self-energy is neglected and only the first line of Eq.~(\ref{eq:Lxph}) is taken into account. 
Hence, the solution of the single-channel approximation of the DMF$^2$RG is equivalent to the RPA with the DMFT vertex in that given channel. 
We have selected the particle-hole crossed channel as a concrete example.
A similar equivalence between single-channel DMF$^2$RG and RPA also holds for the particle-particle and the direct particle-hole channels.


\section{Results}

\begin{figure*}
\includegraphics[width=0.85\textwidth]{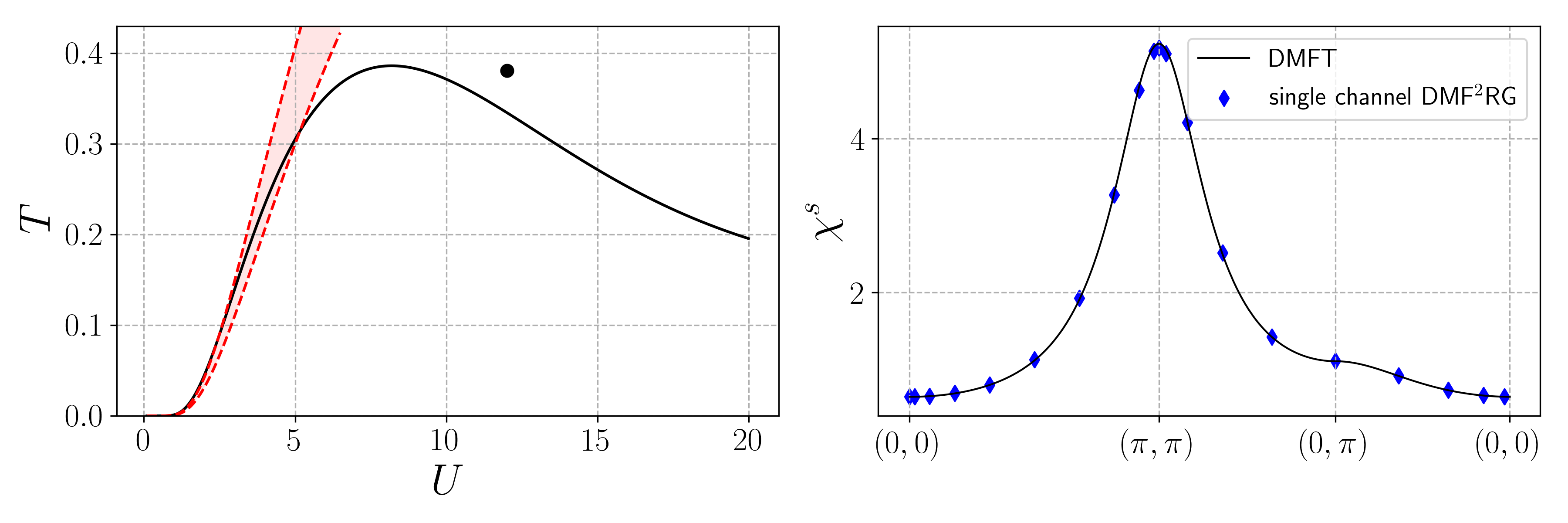}
 \caption{Left panel: DMFT N\'eel temperature as a function of $U$ (black line) for $n=1$ and $t'=0$. The shadowed area depicts the range of transition temperatures obtained from a simplified parametrization of the vertex with a single bosonic frequency variable in each channel. 
 Right panel: Spin susceptibility $\chi^s$ along a path in the BZ zone as computed from RPA with DMFT vertex and self-energy (black solid line), and by the single-channel DMF$^2$RG (blue symbols). Here $U=12t$ and $T=0.038t$, corresponding to the black dot in the left panel.}
\label{fig:sketch} 
\end{figure*}
We will now discuss our results obtained by means of DMF$^2$RG in its full frequency dependent implementation. 
In the first part of this section we test the method at half-filling for both weak and strong interactions. 
We will show that the DMF$^2$RG is able to access the strong coupling regime, 
once the vertex frequency dependence is properly taken into account. 
The second part of the section is dedicated to the more interesting parameter regime away from half-filling, relevant for high temperature superconductivity in cuprates. We will focus on the interplay between the two key players in this regime, 
strong magnetic fluctuations and emerging $d$-wave pairing fluctuations. 

Numerical details are described in Appendix \ref{sec:Numerics}.
The spin susceptibility $\chi_q^s$ with $q=(\bs{q},\Omega)$ is obtained from the two-particle vertex as
\begin{equation}
 \chi_q^s = \int_k \chi_q^0(k)
 + \int_{k,k'} \chi_q^0(k) \, V(k,k'+q,k') \, \chi_q^0(k') ,
\end{equation}
where $\chi_q^0(k) = - G(k) G(k+q)$.
We set $t=1$ in all plots of quantities with dimension energy.


\subsection{Particle-hole symmetric case} 

In this section we focus on the special case of pure nearest neighbor hopping ($t'=0$) at half-filling ($n=1$), where particle-hole symmetry leads to several simplifications.
Due to perfect nesting, the physics is dominated by magnetic fluctuations peaked at $(\pi,\pi)$ for any coupling strength $U$.
We will present results for the magnetic properties of the half-filled 2D Hubbard model, and show that taking the full frequency dependence of the vertex into account is crucial at strong coupling.

In the left panel of Fig.~\ref{fig:sketch}, we show the $U$-dependence of the N\'eel temperature as obtained from the DMFT. The smooth curve is a fit to data points obtained previously by Kunes, \cite{Kunes2011} which are consistent with our own calculations.
We have checked numerically that the N\'eel temperature predicted by the single-channel DMF$^2$RG described in Sec.~\ref{sec:1ch} indeed agrees with the N\'eel temperature computed from the RPA susceptibility with the local DMFT vertex.
The red shadowed area, instead, shows the N\'eel temperature as obtained from the single-channel DMF$^2$RG with an approximate ansatz for the frequency dependence, where only the bosonic frequency dependence of the magnetic fluctuation term $\mathcal{M}^\Lambda$ is taken into account, while the two fermionic frequencies are projected to some arbitrary value. \cite{Taranto2014,Vilardi2017}
Different choices for the projection lead to different estimates for the transition temperature -- hence the shadowed area instead of a single transition line.
As the interaction is increased the difference between the upper and the lower transition 
temperatures increases, reflecting the fact that the quality of the single-frequency approximation deteriorates. As a matter of fact, the error is sizable already for intermediate coupling. 
Eventually, the approximation fails to reproduce the maximum of the N\'eel temperature 
as a function of $U$ and its decrease at large $U$.

On the other hand, we have verified numerically that the single-channel DMF$^2$RG with full frequency dependence reproduces exactly the DMFT results, where the susceptibility is computed from a RPA (ladder sum) with the DMFT vertex. While this agreement is dictated by the analytic proof in Sec.~\ref{sec:1ch}, it is still challenging to reproduce in a numerical evaluation.
To demonstrate the accuracy of the agreement, and thus the performance of our code, we plot the susceptibility along a specific momentum path in the Brillouin zone computed with both methods (right panel of Fig.~\ref{fig:sketch}), for a parameter set at strong coupling where the single-frequency approximation fails drastically. 

\begin{figure*}
\centering
\includegraphics[width=0.98\textwidth]{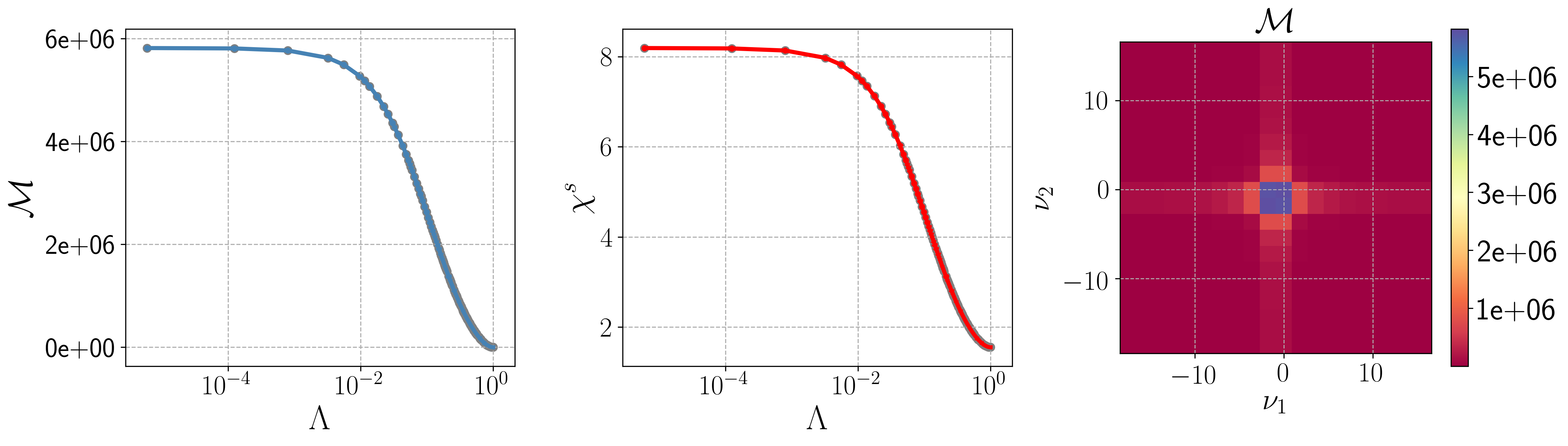}
\caption{Left panel: Flow of the maximum of the magnetic fluctuation term as function of 
 the flow parameter $\Lambda$. 
 Center panel: Flow of the magnetic susceptibility at $\bs{Q}=(\pi,\pi)$ and $\Omega=0$. 
 Right panel: Frequency dependence of the magnetic fluctuation term for momentum
 $\bs{Q}=(\pi,\pi)$ and vanishing bosonic frequency $\Omega=0$. 
 Parameters: $U=16t$, $T=0.29t$, $t'=0$ and $n=1$.}
\label{fig:magnetic}
\end{figure*}
 
The decrease of the N\'eel temperature at large $U$ is known to be associated with a change in the mechanism leading to an antiferromagnetic ground state, from Slater-type to Heisenberg-type. \cite{Toschi2005,Sangiovanni2006,Taranto2012}
The failure of the single-frequency approximation in the intermediate to strong coupling regions reveals that the vertex acquires a frequency structure that cannot be reproduced by a single bosonic frequency only.

We now turn to the first complete DMF$^2$RG calculation at strong coupling. 
Here the flow of the vertex is computed with all the channels taken into account. 
In the particle-hole symmetric case, the DMF$^2$RG always exhibits an antiferromagnetic instability toward a N\'eel state at low temperature. 
In Fig.~\ref{fig:magnetic} we show, from left to right, the flow of the maximum of the magnetic fluctuation term, the flow of the maximum of the magnetic susceptibility, 
and the magnetic fluctuation term for $\bs{Q} = (\pi,\pi)$ and $\Omega = 0$ at the end of the flow, as a function of the fermion frequencies. The coupling strength is $U=16t$, and the temperature $T=0.29t$ slightly above the N\'eel temperature.
We see that DMF$^2$RG is able to recover convergent results at strong coupling, where the conventional fRG is clearly inapplicable. 
Note that the vertex maximum at strong coupling can be thousands or even millions of times larger than the hopping, as can be seen from the left panel of Fig.~\ref{fig:magnetic}. 
However, the maximum is very sharp in frequency space -- see the right panel of 
Fig.~\ref{fig:magnetic}. This, together with the self-energy, leads to relatively moderate values of the magnetic susceptibility shown in the central panel of Fig.~\ref{fig:magnetic}. 

In weak coupling fRG calculations \cite{Metzner2012} the flow is usually stopped when the largest vertex component exceeds a certain value $V_{\mathrm{max}}$ of the order of ten or hundred times the hopping, since this is typically a precursor of a divergence, accompanied by a divergence of a susceptibility, and the weak coupling truncation is at least questionable at this point.
At strong coupling, we see that the magnetic fluctuation contribution to the two-particle vertex can be huge in a small frequency regime, while the magnetic susceptibility is only moderately enhanced at the chosen temperature, and the flow remains stable.
At weak coupling, the dependence of the vertex on the fermion frequencies is much more shallow. \cite{Vilardi2017} 

The instability criterion in conventional fRG, suggested by weak coupling arguments \cite{Salmhofer2001} and based on the size of the two-particle vertex, is thus misleading at strong coupling. In fact, at strong coupling already the DMFT vertex can be very large for certain frequencies, while the susceptibility, which contains a summation over the fermionic frequencies of the vertex, can still be moderate. 
Hence, rather than looking at the maximal value of the vertex, the instabibility criterion should be defined by the maximum of the corresponding susceptibility.

\begin{figure}
\centering
\includegraphics[width=0.48\textwidth]{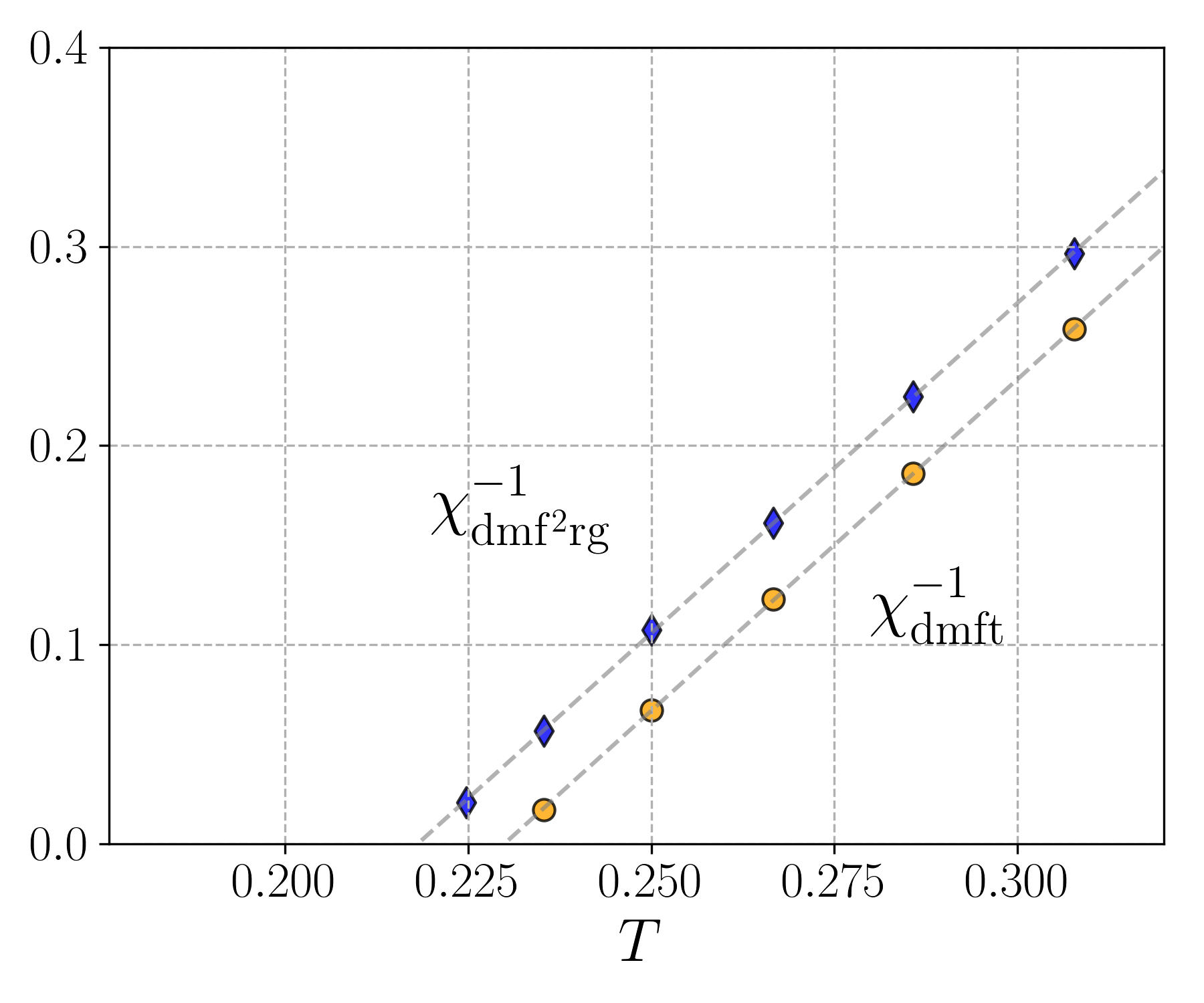}
\caption{Inverse of the static magnetic susceptibility for $\bs{Q} = (\pi,\pi)$ as a function of the temperature for $U=4t$ in DMF$^2$RG and in RPA with DMFT vertices for $n=1$ and $t'=0$.} 
\label{fig:inv_susc} 
\end{figure} 

In Fig.~\ref{fig:inv_susc} we plot the inverse of the magnetic susceptibility for $\Omega=0$ and $\bs{Q} = (\pi,\pi)$, at an intermediate coupling as a function of the temperature.
An extrapolation of $(\chi^s)^{-1}$ indicates a finite N\'eel temperature.
For a comparison we also show the same quantity as computed by the RPA with DMFT vertices. 
One can see that the N\'eel temperature in DMF$^2$RG is only slightly reduced 
compared to the DMFT results, which, in turn, is much smaller than the temperature predicted by the standard RPA.
In conventional fRG, fluctuations in the non-magnetic channels (mostly pairing) substantially reduce the N\'eel temperature. On the local level, these effects are already taken into account by the DMFT, while a further reduction of the N\'eel temperature due to non-local fluctuations in the non-magnetic channels turns out be to less pronounced. 

At half filling and with $t'=0$, a divergent spin susceptibility signaling a magnetic instabibility at low temperature is found in our calculations for any coupling strength.
However, an ordered magnetic state breaking the $SU(2)$ spin symmetry is excluded at finite temperature in two dimensions by the Mermin-Wagner theorem. \cite{Mermin1966}
The truncation of non-local fluctuation contributions underlying our present implementation of the DMF$^2$RG misses the order parameter fluctuations preventing the magnetic order at finite temperatures. This deficency could be cured by including thermal order parameter fluctuations using the techniques developed by Baier et al.~\cite{Baier2004} for the plain fRG.


\subsection{Finite doping}

Let us now switch to the finite doping case in a parameter range relevant for cuprates.
The ratio of next-to-nearest neighbor hopping and nearest neighbor hopping is $t'/t = -0.2$ in the entire section.

\begin{figure}
\includegraphics[width=0.48\textwidth]{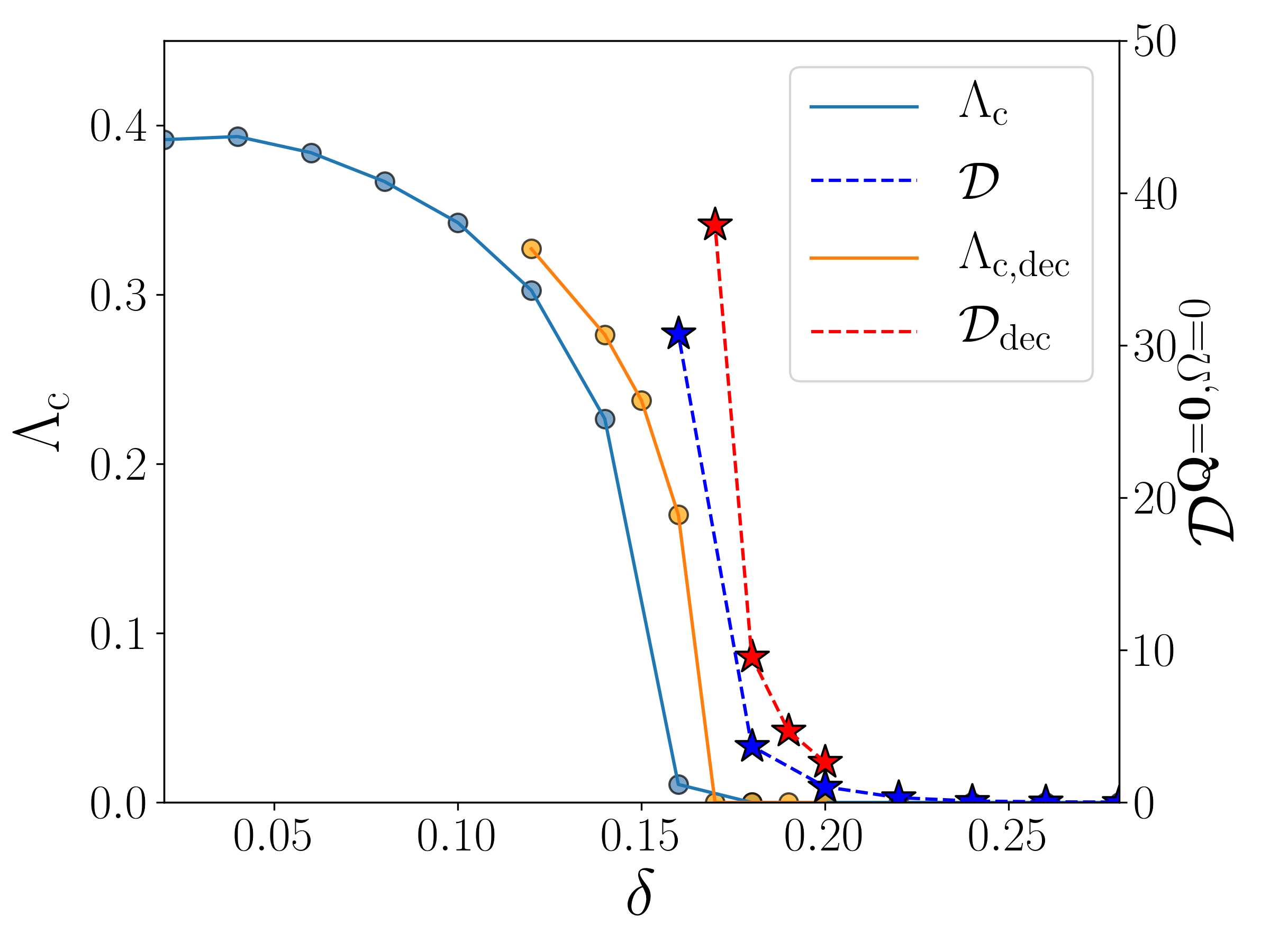}
\caption{Left axis: Critical flow parameter $\Lambda_{\rm c}$ for the antiferromagnetic
 instability as a function of doping $\delta=1-n$ in full DMF$^2$RG (blue circles) and in single-channel DMF$^2$RG (orange circles), respectively.
 Right axis: Maximum of the $d$-wave pairing interaction $\mathcal{D}$ from the full DMF$^2$RG (blue stars) and in a decoupling approximation (red stars), respectively.
 The lines connecting the symbols are guides to the eye.
 Parameters are: $U=8t$, $T=0.08t$ and $t'=-0.2t$. } 
\label{fig:dop_phase_diag}
\end{figure}


\subsubsection{Magnetic fluctuations}

In Fig.~\ref{fig:dop_phase_diag} we show the critical flow parameter $\Lambda_{\rm c}$ as a function of doping for $U=8t$ and $T=0.08t$. Assuming a hopping value for cuprates of $t \approx 0.4$eV, the chosen temperature is thus about 350K. 
We observe a magnetic instability for all dopings smaller than $\delta_{\mathrm{c}}=0.18$. 
For higher doping values the flow reaches $\Lambda=0$ without encountering any instability.
Decreasing the temperature to $T=0.044t$, we only observe a very slight increase of 
the critical doping value. Hence, from our results, we see that the critical doping for a magnetic instabibility $\delta_\mathrm{c}$ remains about 0.18 down to the lowest temperatures. This value is roughly comparable with the maximal doping range for which  the pseudogap is experimentally observed. \cite{Taillefer2010}
Hence, the large magnetic fluctuations leading to the instability of the flow should not be associated with spontaneous symmetry breaking, but rather with the onset of the pseudogap. 
The instability occurs at the commensurate antiferromagnetic wave vector $(\pi,\pi)$ for $\delta<0.16$, and at incommensurate wave vectors of the form $(\pi - 2\pi\eta,\pi)$ with $\eta > 0$ for larger values of the doping.
These results are in line with a similar transition from commensurate to incommensurate magnetic fluctuations revealed by the DMFT-RPA susceptibility (with DMFT self-energy and vertex corrections). \cite{Vilardi2018}

\begin{figure}
\includegraphics[width=0.5\textwidth]{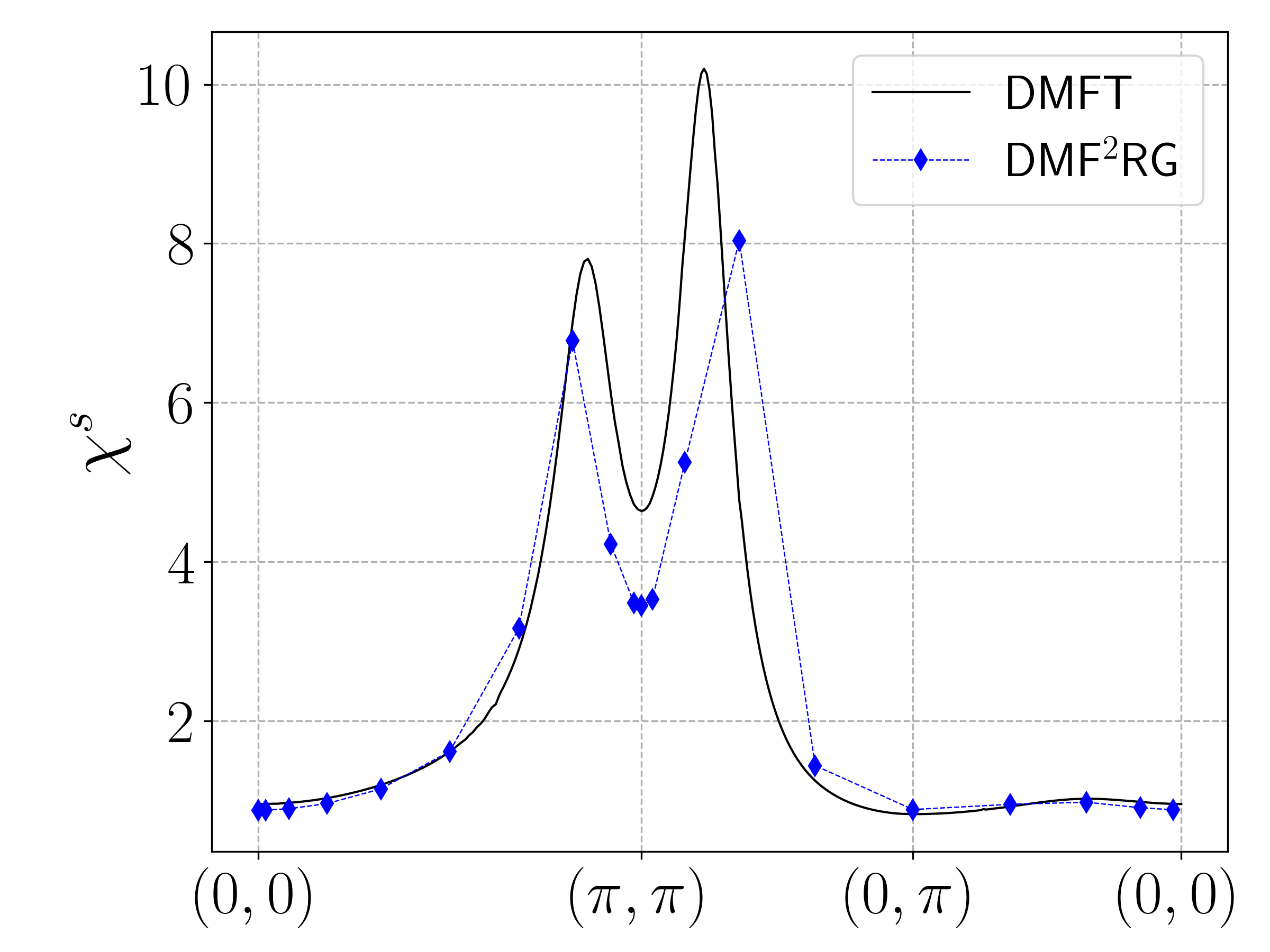}
\caption{Static magnetic susceptibility in DMFT-RPA (black line) and in full
 DMF$^2$RG (blue points) along a specific path in the BZ.
 Parameters: $U=8t$, $T=0.08t$, $t'=-0.2t$ and $\delta=0.18$.}
\label{fig:mag_susc}
\end{figure} 
In Fig.~\ref{fig:mag_susc} we compare the magnetic susceptibility of DMF$^2$RG with the 
one from RPA with DMFT vertex for doping $\delta=0.18$ along a specific path in the BZ.
The two susceptibilities are qualitatively similar, showing that the inclusion of the non-magnetic channels leads only to minor quantitative modifications in this parameter regime. In particular we observe that in both cases $(\pi,\pi)$ is a marked local minimum. The maximum of the susceptibility in DMF$^2$RG seems to be shifted to a slightly 
different incommensurate wave vector compared to the DMFT-RPA, but the limited momentum resolution of the DMF$^2$RG calculation does not allow for a conclusive statement. 
\begin{figure}
\includegraphics[width=0.49\textwidth]{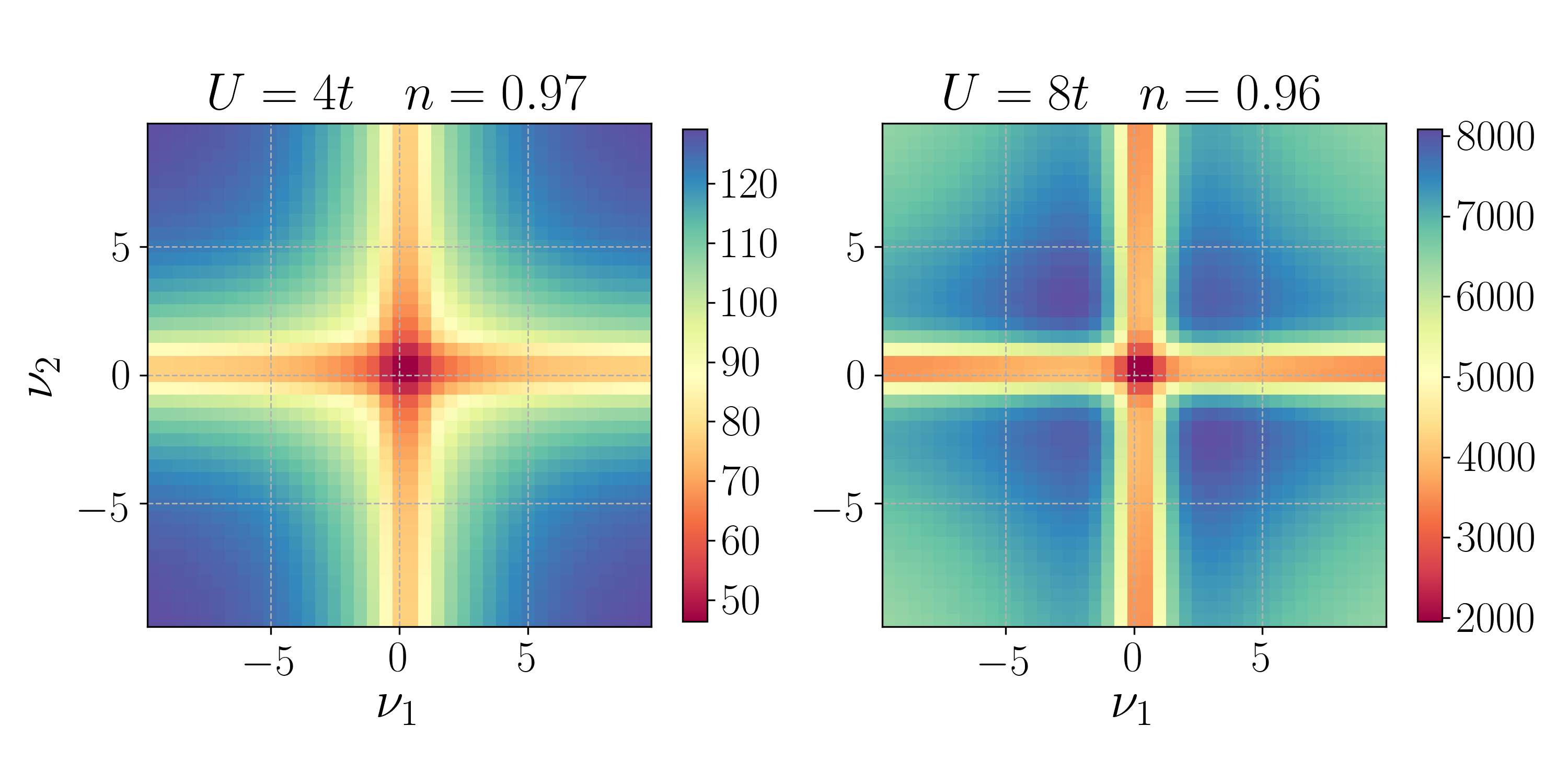}
\caption{Frequency dependence of the magnetic fluctuation channel at weak (left) and
 strong (right) coupling close to half-filling for $T=0.08t$ and $t'=-0.2t$.}
\label{fig:mag_freq} 
\end{figure}

To highlight the different frequency structures that arise in different coupling regimes, we show in Fig.~\ref{fig:mag_freq} the frequency dependence of $\mathcal{M}^\Lambda$ for $\Omega=0$ and $\bs{Q}=(\pi,\pi)$ at moderate and strong coupling, with $\Lambda$ slightly below the critical value $\Lambda_c$. 
At moderate coupling ($U=4t$) the maximal value of $\mathcal{M}^\Lambda$ is observed for asymptotically large values of $\nu_1$ and $\nu_2$ in the frequency region where 
the channel competition is less effective. The cross shaped structure, that can be 
ascribed to the effect of the feedback from the other channels, \cite{Wentzell2016} 
on the other hand, decreases the value of $\mathcal{M}^\Lambda$. At strong coupling ($U=8t$), the cross shaped structure is still decreasing $\mathcal{M}^\Lambda$, but the maximal values are not in the asymptotic region, but in a localized area for limited values of $\nu_1$ and $\nu_2$ (and away from the cross shaped structure). Although a complete explanation of these features in Matsubara frequency space is complicated, 
they hint to a different nature of the magnetic fluctuations at weak or moderate and at strong coupling.


\subsubsection{Self-energy}

\begin{figure}
\includegraphics[width=0.45 \textwidth]{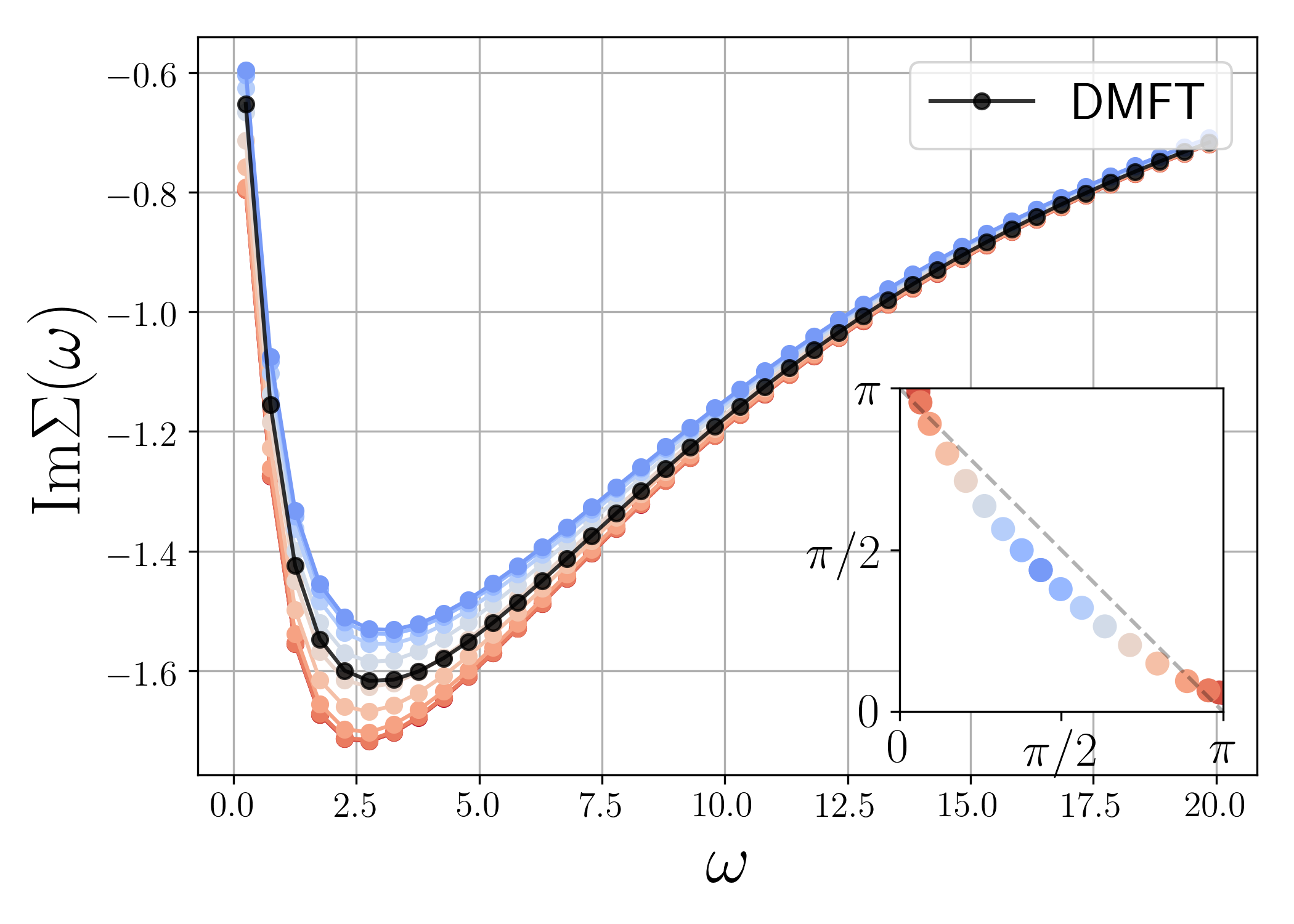}
\caption{Imaginary part of the self-energy as a function of Matsubara frequency 
 for different points in the BZ along the noninteracting Fermi surface (see inset). 
 The local DMFT self-energy is shown in black. 
 Parameters: $U=8t$, $T=0.08t$, $t'=-0.2t$ and $\delta=0.18$.} 
\label{fig:self_energy}
\end{figure} 
In Fig.~\ref{fig:self_energy} we show the imaginary part of the self-energy in Matsubara space for different points in the BZ, and for $\delta=0.18$. For this doping value, the flow reaches the final $\Lambda$ without encountering any instability, but the magnetic fluctuations are already strongly enhanced. 
Therefore one could have expected some signature of a strong momentum differentiation in the self-energy, associated with a suppression of the spectral weight in the antinodal region. This is not observed in our calculation. The self-energy obtained from the DMF$^2$RG does not deviate qualitatively from the DMFT result, and exhibits only a slight decrease of the quasiparticle weight \cite{footnote1} close to the antinodes. This result is very similar to the one we obtained at weak coupling within a conventional fRG scheme with full-frequency dependence. \cite{Vilardi2017}


\subsubsection{$d$-wave pairing fluctuations}

As discussed above, the pairing and density channels do not strongly affect the magnetic one. However the reverse is not true: the magnetic channel generates $d$-wave pairing fluctuations which, for lower temperatures, are expected to give rise to a pairing instability.

In Fig.~\ref{fig:dop_phase_diag} (see stars and right axis) we show the maximal value of $\mathcal{D}^\Lambda$ for the lowest accessible value of $\Lambda$, which measures the strength of the $d$-wave pairing interaction. For dopings much larger than $\delta_\mathrm{c}$ the pairing interaction is very small. Decreasing the doping from $0.2$ to $0.16$ the $d$-wave pairing interaction rapidly increases. Decreasing the doping further, the flow runs into the magnetic instability and has to be stopped at the critical flow parameter $\Lambda_\mathrm{c}$. The $d$-wave interaction at the critical scale $\Lambda_c$ then drops again, to very small values.

These results can be interpreted as follows. For $\delta \gtrsim \delta_\mathrm{c}$ the magnetic fluctuations become very strong and the large magnetic channel drives the $d$-wave interaction to large values. When the doping is decreased further, the flow has to be stopped before the $d$-wave interaction can fully develop.
In the context of the conventional fRG it has been frequently observed \cite{Halboth2000,Eberlein2014} that the $d$-wave pairing increases quite rapidly at a late stage of the flow, as compared to the more gradual increase of the magnetic channel, which sets in already at high energy scales. 
While the flow parameter in DMF$^2$RG is a measure of non-locality rather than an energy scale, the retarded but then rapid formation of pairing interactions seems to be typical here, too.


To confirm the magnetic pairing mechanism, in Fig.~\ref{fig:dop_phase_diag} we also present the critical value $\Lambda_{\rm c}$ and the pairing interaction $\mathcal{D}^\Lambda$ within a simplified approximation, where we neglect the flow of the self-energy and we set $\mathcal{C}^{\Lambda} = \mathcal{S}^{\Lambda} = 0$, while the magnetic channel is treated at the single-channel level as in Sec.~\ref{sec:1ch}.
As a consequence, the $d$-wave pairing channel receives contributions only from the magnetic channel and the pairing channel itself.
In this approach the feedback of charge and $s$-wave pairing channels is taken into account only at the DMFT level. The $d$-wave pairing channel does not receive any contribution from the DMFT vertex, since the latter is local. 
The resulting critical flow parameter $\Lambda_{\rm c}$, shown in orange in Fig.~\ref{fig:dop_phase_diag}, is always slightly larger than the one from the full DMF$^2$RG. This confirms that the channel competition has only a modest detrimental effect on the magnetic fluctuations.
The maximal doping value for which the magnetic instability is observed increases. 
A sizable $d$-wave pairing interaction sets in for higher values of the doping, too. 
There is no major difference in the $d$-wave pairing interaction compared to the full DMF$^2$RG where all the channels are included, supporting the hypothesis that $d$-wave pairing is mostly driven by the nonlocal magnetic channel. 

\begin{figure}
\includegraphics[width=0.48\textwidth]{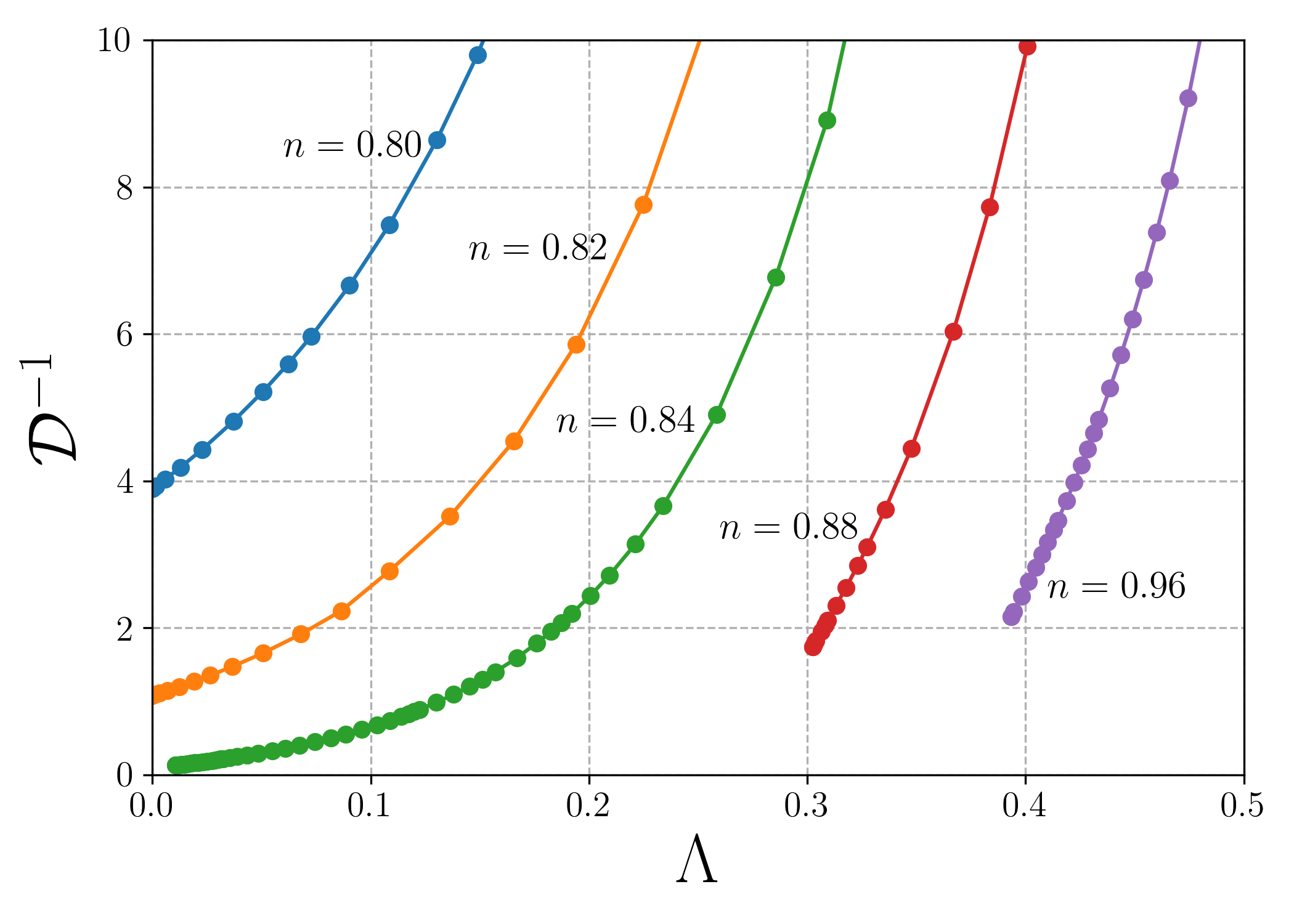}
\caption{Inverse $d$-wave channel as a function of flow parameter $\Lambda$ for various fillings. Parameters: $U=8t$, $T=0.08t$ and $t'=-0.2$. } 
\label{fig:dwave_inverse}
\end{figure} 
\begin{figure}
\includegraphics[width=0.48\textwidth]{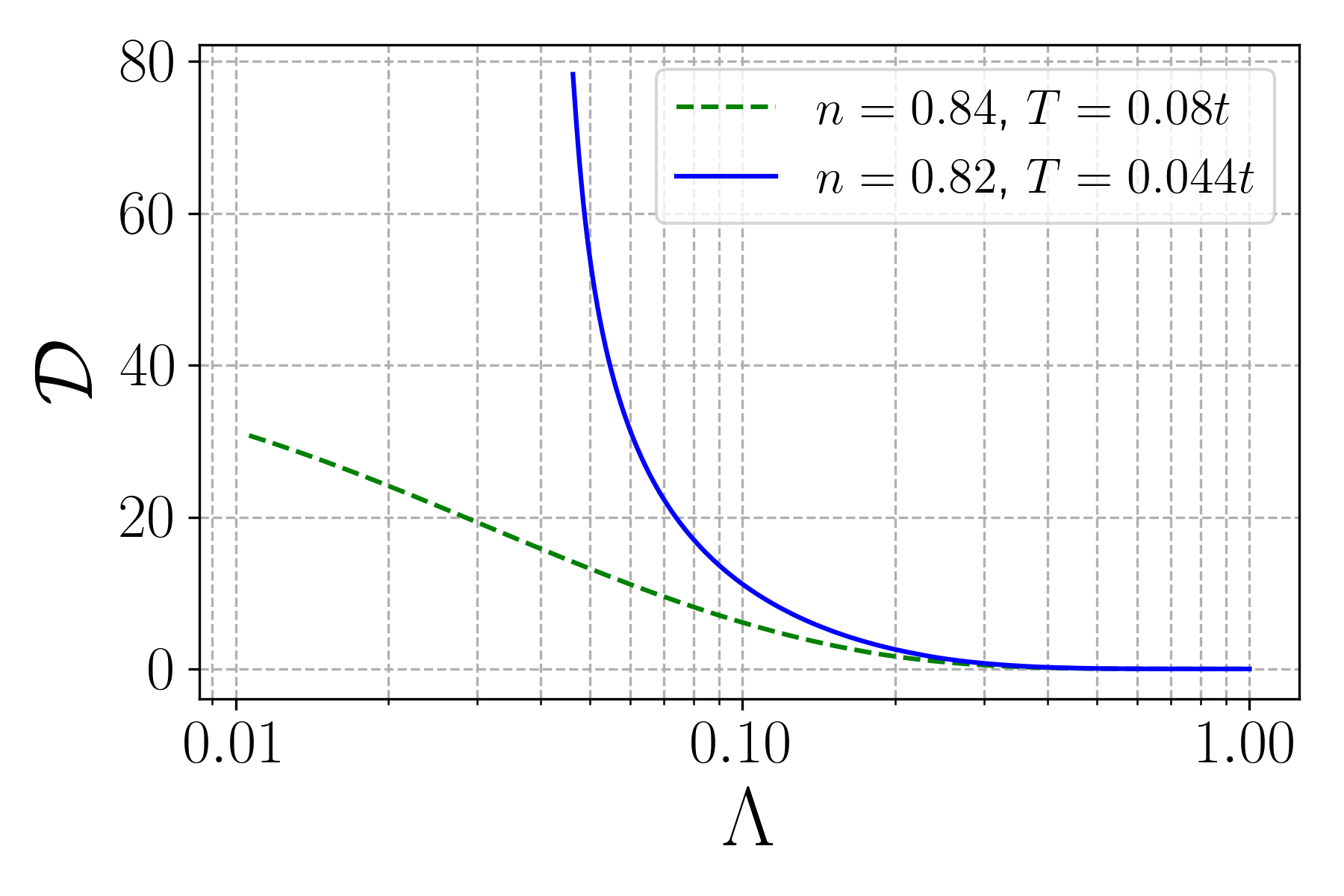}
\caption{Flow of the $d$-wave pairing channel at higher and lower temperature for $U=8t$ and $t'=-0.2$.} 
\label{fig:dwave_flow}
\end{figure} 

In Fig.~\ref{fig:dwave_inverse} we show the inverse $d$-wave pairing interaction
$\mathcal{D}^{-1}$ for $\bs{Q}=(0,0)$ and $\Omega=0$, as a function of the flow parameter $\Lambda$ for different fillings. The parameters are the same as in Fig.~\ref{fig:dop_phase_diag}. 
For $n=0.88$ and $n=0.96$, the flow is shown up to the critical value $\Lambda_{\rm c}$ at which the magnetic instability occurs. 
Approaching half-filling $n=1$, the $d$-wave pairing correlations increase but cannot develop further due to the magnetic instabibility which prevents a continuation of the flow to smaller $\Lambda$.

Finally, let us discuss the role of the temperature. 
The results discussed so far are for a temperature $T=0.08t$, roughly comparable to room temperature and thus much higher than the maximal temperatures for which $d$-wave 
superconductivity has been observed. Therefore, we do not expect a $d$-wave pairing instability at this temperature, but the onset of a large $d$-wave pairing interaction is likely a high-temperature precursor of a superconducting phase at lower temperature. 

Different theoretical studies yield different estimates for the maximal temperature for 
which superconductivity is observed for the Hubbard model on the square lattice. For example, while cluster extensions of the DMFT \cite{Chen2015,Fratino2016} find a higher scale of $T \approx 0.03t$, diagrammatic methods \cite{Kitatani2018,Vuciceciv2017} yield superconductivity only for temperatures below $T \approx 0.01t$. Experimentally, the maximal superconducting temperature observed for cuprates is $\mathcal{O}(100)$K, which roughly corresponds to $T \sim 0.02t$ in units of the nearest-neigbhor hopping amplitude. Hence, we expect that we need to decrease the temperature at least by a factor of two compared to what we have achieved so far, to observe a pairing instabibility. 

Due to the high computational cost of low-$T$ calculations, we cannot reach the superconducting transition temperature at the moment.
However, to better understand the evolution of the $d$-wave fluctuations at lower temperatures, we have performed few computations at a reduced (compared to the above) temperature $T=0.044t$. A result is shown in Fig.~\ref{fig:dwave_flow}, where we show the flow of the maximum of the $d$-wave pairing channel $\mathcal{D}$ for the doping value for which the $d$-wave pairing is most pronounced. A flow at $T=0.08t$ for a slightly different filling is also shown for comparison.
Our expectation is that, as the temperature is further decreased, the relative relevance of the $d$-wave pairing should increase and its flow become more steep, until, eventually  the $d$-wave pairing becomes larger than the magnetic one. This is indicated by the comparison in Fig.~\ref{fig:dwave_flow}, where the pairing interaction at the lower temperature is not only much larger, but also has a larger slope.
In both cases the critical value $\Lambda_{\rm c}$ is set by the instability in the magnetic channel, but the $d$-wave pairing interaction is much larger for the lower temperature. 

All these observations lead us to the conclusion that also in the strong-coupling regime the magnetic fluctuations can generate large $d$-wave pairing interactions leading ultimately a pairing instability at sufficiently low temperatures. 

%
%

\section{Conclusion}

In summary, we have developed the DMF$^2$RG, a combination of DMFT and fRG proposed several years ago by Taranto et al. \cite{Taranto2014}, to a practical method that allows for the computation of local and non-local dynamical correlation functions in strongly interacting lattice fermion systems.
The fRG flow starts from the DMFT solution for the self-energy and the two-particle vertex. Local correlations are treated non-perturbatively by the DMFT, while the flow is driven exclusively by non-local correlations, such that only the latter are affected by the truncation of the exact flow equation hierarchy. This improves the accuracy of the method substantially compared to the early version of the DMF$^2$RG.

Another crucial improvement concerns the frequency dependence of the two-particle vertex. We have shown that a reduction to a single bosonic frequency variable in each fluctuation channel (magnetic, charge, and pairing) is inadequate already at moderate coupling, and fails completely in the strong coupling regime. Hence, we have taken the full frequency dependence into account. While this is challenging due to the large number of variables and the extremely singular frequency dependence of the vertex at strong coupling, we have managed to perform calculations with a sufficiently large number of frequencies down to temperatures of about one percent of the band width.

We have applied our implementation of the DMF$^2$RG to the two-dimensional Hubbard model with interactions up to $U=8t$, both at half-filling and in the hole-doped regime. Most of the calculations were performed at a fixed temperature $T=0.08t$.
Magnetic correlations dominate from half-filling up to 18 percent hole doping. In that regime the magnetic fluctuation term diverges for a certain critical flow parameter $\Lambda_c$, signaling an antiferromagnetic instabibility. The magnetic fluctuations are peaked at the N\'eel wave vector $(\pi,\pi)$ for doping $\delta < 0.16$, and at incommensurate wave vectors of the form $(\pi-2\pi\eta,\pi)$ for larger doping. Their strength is only mildly reduced by non-local fluctuations in other (non-magnetic) channels, while a substantial reduction from local correlations is already taken into account by the DMFT. The antiferromagnetic instability obtained at finite temperature is due to missing feedback of thermal order parameter fluctuations in our truncation of the fRG flow (see below). Hence, the divergence of the magnetic channel in our calculation should rather be associated with the pseudogap formation rather than magnetic long-range order.

At the edge of the regime dominated by magnetic fluctuations, near $\delta = 0.18$, we find sizable $d$-wave pairing fluctuations. Lowering the temperature to $T=0.044t$ they almost diverge. Hence, the system is not far from a pairing instabibility at that temperature, consistent with superconductivity in the temperature range observed for cuprates. Switching off the non-magnetic fluctuation channels we could show that the dominant driving mechanism for $d$-wave pairing is magnetic, as suggested by various physical arguments, \cite{Scalapino2012} and confirmed for moderate interactions by plain fRG calculations. \cite{Metzner2012} 

The divergence of the magnetic fluctuation term and the magnetic susceptibility in the doping range from half-filling to $\delta=0.18$ indicates the importance of non-local magnetic correlations in that regime, but also a breakdown of our truncation at the critical flow parameter $\Lambda_c$. The Mermin-Wagner theorem excludes magnetic long-range order at any finite temperature, and the magnetic susceptibility should diverge only in the zero temperature limit. Magnetic order is prevented by thermal order parameter fluctuations via a destructive feedback mechanism that is not captured by our truncation of the flow equations. The most efficient way of dealing with these effects is by introducing a bosonic order parameter field via a Hubbard-Stratonovich decoupling of the dominant magnetic interactions. A relatively simple truncation of the fRG flow for the order parameter fluctuations then pushes the magnetic phase transition to zero temperature, in agreement with the Mermin-Wagner theorem. \cite{Baier2004}
Alternatively, one may also recover the Mermin-Wagner theorem in a purely fermionic flow via the recently developed multi-loop truncation of the flow equation hierarchy. \cite{Kugler2018,Tagliavini2018}
Extending such refinements to the DMF$^2$RG is one of the most promising future directions. The temperature range in which the present implementation of the DMF$^2$RG breaks down due to the divergent magnetic fluctuation term would then become accessible. We expect strong but finite magnetic correlations in that regime, and thus a fertile soil for pseudogap behavior and $d$-wave pairing.


\begin{acknowledgments} 

We are grateful to S.~Andergassen, A.~Eberlein, P.~Hansmann, C.~Hille, A.~Tagliavini, and A.~Toschi for useful discussions. 

\end{acknowledgments}


\newpage

\begin{appendix}


\section{Flow equations}
\label{sec:FlowEquations}

The flow equation for the magnetic channel has been presented in Sec.~\ref{sec:vertex}.
Here we present the expressions for the flow equations in the pairing and in the charge channels. The flow equation for the $s$-wave pairing channel reads
\begin{widetext}
\begin{equation}
 \frac{d}{d\Lambda} \mathcal{S}^{\Lambda}_{\bs{Q},\Omega}(\nu_1,\nu_3) = 
  T \sum_\nu L^{\mathrm{s},\Lambda}_{\mathbf{Q},\Omega}(\nu_1,\nu) P^{\mathrm{s},\Lambda}_{\bs{Q},\Omega}(\nu)
  L^{\mathrm{s},\Lambda}_{\mathbf{Q},\Omega}(\nu,\nu_3),
\end{equation}
with
\begin{equation}
 P^{\mathrm{s},\Lambda}_{\bs{Q},\Omega}(\omega) = \int_{\bs{p}}
 G^{\Lambda}(\bs{p},\omega)S^{\Lambda}(\bs{Q}-\bs{p},\Omega-\omega) +
 G^{\Lambda}(\bs{Q}-\bs{p},\Omega-\omega) S^{\Lambda}(\bs{p},\omega), 
\label{eq:app:P_pp_s}
\end{equation}
and
\begin{eqnarray} 
\label{eq:Lswave}
 L^{\mathrm{s},\Lambda}_{\mathbf{Q},\Omega}(\nu_1,\nu_3) &=&
 V_{\rm dmft}(\nu_1,\Omega-\nu_1,\nu_3)
 -\mathcal{S}^{\Lambda}_{\bs{Q},\Omega} (\nu_1,\nu_3) \nonumber \\
 &+& \int_{\bs{p}} 
 \Big[ \mathcal{M}^{\Lambda}_{\bs{p},\nu_3-\nu_1}(\nu_1,\Omega-\nu_1) 
 + \frac{1}{2} \mathcal{M}^{\Lambda}_{\bs{p},\Omega-\nu_1-\nu_3}(\nu_1,\Omega-\nu_1) -
 \frac{1}{2} \mathcal{C}^{\Lambda}_{\bs{p},\Omega-\nu_1-\nu_3}(\nu_1,\Omega-\nu_1) \Big]. 
\end{eqnarray}
The flow equation for the $d$-wave pairing channel reads
\begin{equation}
 \frac{d}{d\Lambda} \mathcal{D}^{\Lambda}_{\bs{Q},\Omega}(\nu_1,\nu_3) = 
 T \sum_\nu L^{\mathrm{d},\Lambda}_{\bs{Q},\Omega}(\nu_1,\nu)
 P^{\mathrm{d},\Lambda}_{\bs{Q},\Omega(\nu)}
 L^{\mathrm{d},\Lambda}_{\bs{Q},\Omega} (\nu,\nu_3), 
\label{eq:dwaveflow}
\end{equation}
with
\begin{equation}
 P^{\mathrm{d},\Lambda}_{\bs{Q},\Omega}(\omega) =
 \int_{\bs{p}} \left[ f_{\mathrm{d}}\left( \bs{Q}/2 - \bs{p} \right) \right]^2 
 \left[ G^{\Lambda}(\bs{p},\omega)S^{\Lambda}(\bs{Q}-\bs{p},\Omega-\omega) +G^{\Lambda}(\bs{Q}-\bs{p},\Omega-\omega)
S^{\Lambda}(\bs{p},\omega) \right], 
\label{eq:app:P_pp_d}
\end{equation}
and
\begin{eqnarray} 
\label{eq:Ldwave}
 L^{\mathrm{d},\Lambda}_{\bs{Q},\Omega}(\nu_1,\nu_3) &=&
 - \mathcal{D}^{\Lambda}_{\bs{Q},\Omega}(\nu_1,\nu_3)
 + \frac{1}{2}\int_{\bs{p}} \left(\cos{p_x}+\cos{p_y}\right) \nonumber \\
 &\times& \Big[ \mathcal{M}^{\Lambda}_{\bs{p},\nu_3-\nu_1}(\nu_1,\Omega-\nu_1) 
 + \frac{1}{2} \mathcal{M}^{\Lambda}_{\bs{p},\Omega-\nu_1-\nu_3}(\nu_1,\Omega-\nu_1)
 - \frac{1}{2} \mathcal{C}^{\Lambda}_{\bs{p},\Omega-\nu_1-\nu_3}(\nu_1,\Omega-\nu_1) \Big] .
\end{eqnarray}
Since $\mathcal{D}^{\Lambda}$ is generated exclusively by fluctuation contributions (not by the DMFT vertex $V_{\rm dmft}$), see Eq. (\ref{eq:Ldwave}), it is the channel which is most sensitive to approximations on the frequency dependence.  

The flow equation for the charge channel reads
\begin{equation}
 \frac{d}{d\Lambda} \mathcal{C}^{\Lambda}_\bs{Q,\Omega}(\nu_1,\nu_2) = -T\sum_\nu
 L^{\mathrm{c},\Lambda}_{\bs{Q},\Omega}(\nu_1,\nu) P^{\Lambda}_{\bs{Q},\Omega}(\nu) 
 L^{\mathrm{c},\Lambda}_{\bs{Q},\Omega}(\nu,\nu_2-\Omega), 
\end{equation}
with $P^{\Lambda}_{\bs{Q},\Omega}(\omega)$ as in Eq.~(\ref{eq:Pph}), and
\begin{eqnarray}  
 L^{\mathrm{c},\Lambda}_{\bs{Q},\Omega}(\nu_1,\nu_2) &=&
 2 V_{\rm dmft}(\nu_1,\nu_2,\Omega+\nu_1) - V_{\rm dmft}(\nu_2,\nu_1,\Omega+\nu_1)
 - \mathcal{C}^{\Lambda}_{\bs{Q},\Omega}(\nu_1,\nu_2)
 \nonumber \\ 
 &+& \int_{\bs{p}} \Big [
 - 2 \mathcal{S}^{\Lambda}_{\bs{p},\nu_1+\nu_2}(\nu_1,\nu_2-\Omega) + \mathcal{S}^{\Lambda}_{\bs{p},\nu_1+\nu_2}(\nu_1,\Omega+\nu_1)
 + \frac{3}{2} \mathcal{M}^{\Lambda}_{\bs{p},\nu_2-\nu_1-\Omega}(\nu_1,\nu_2)
 + \frac{1}{2} \mathcal{C}^{\Lambda}_{\bs{p},\nu_2-\nu_1-\Omega}(\nu_1,\nu_2)
 \nonumber \\ 
 &&+ (\cos Q_x + \cos Q_y) \Big( \mathcal{D}^{\Lambda}_{\bs{p},\nu_1+\nu_2}(\nu_1,\nu_2-\Omega) -\frac{1}{2} \mathcal{D}^{\Lambda}_{\bs{p},\nu_1+\nu_2}(\nu_1,\Omega+\nu_1) \Big) \Big] .
 \label{eq:Lc}
\end{eqnarray}
\end{widetext}

The form factor decomposition allows to decouple the momentum integrals, in the calculation of the $L$'s, Eqs.~(\ref{eq:Lxph}), (\ref{eq:Lswave}), (\ref{eq:Ldwave}) and (\ref{eq:Lc}), from the frequency summations in the flow equations, hence reducing the numerical effort.   


\section{Numerical details}
\label{sec:Numerics}

We first compute the DMFT loop and the DMFT vertex function with an exact diagonalization (ED) method \cite{Caffarel1994} by discretizing the conduction electron bath of the AIM with 4 sites.
The DMFT vertex is computed in a box containing the first 80 positive and the first 80 negative Matsubara frequencies for each of the three frequency variables, while outside the box we extrapolate the frequency dependence with asymptotic functions as described in Ref.~\onlinecite{Wentzell2016}. 
The numerical setup of the flow equations is similar to Ref.~\onlinecite{Vilardi2017}. 
We use different patching schemes in momentum space for the self-energy and for the vertex. 
We use 29 patches for the bosonic momentum dependence ($\bs{Q}$) of the vertex with more accuracy in the corners around $(0,0)$ and $(\pi,\pi)$, where we expect the instability vectors. For the momentum dependence of the self-energy we use 44 patches adapted to the shape of the noninteracting Fermi surface. 
For the frequency dependence of the vertex, we rewrite $\mathcal{S}$, $\mathcal{D}$, 
$\mathcal{C}$ and $\mathcal{M}$ as functions of three bosonic frequency variables.
For each variable the first 40 positive and first 40 negative Matsubara frequencies are kept explicitly, while beyond we extrapolate the asymptotic behaviour. The bosonic representation of the frequency dependence simplifies the numerical treatment of the asymptotic behaviour. 

For the doped case, we keep the filling fixed during the flow by properly adjusting an 
additive constant in the real part of the self-energy. 

\end{appendix}


\begin{thebibliography}{2}
\bibitem{footnote1} The quasiparticle weight can be inferred from the slope
 of the self-energy at the first few Matsubara frequencies.
\end{thebibliography}
\end{document}